\documentclass[fleqn,usenatbib,twocolumn]{mnras}
\usepackage{pdflscape}
\usepackage{newtxtext,newtxmath}
\usepackage{bm}
\usepackage{multirow}

\usepackage[T1]{fontenc}

\usepackage{comment}
\usepackage{enumerate}

\usepackage{longtable}

\DeclareRobustCommand{\VAN}[3]{#2}
\let\VANthebibliography\thebibliography
\def\thebibliography{\DeclareRobustCommand{\VAN}[3]{##3}\VANthebibliography}

\usepackage{graphicx}	
\usepackage{amsmath}	

\title[The distribution and redshift dependence of ${\rm DM_{host}}$]{Unveiling the distribution and redshift dependence of host galaxy dispersion measures using localized fast radio bursts}

\author[Y. Sang and H.-N. Lin]{
Yu Sang$^{1}$,
Hai-Nan Lin$^{2}$\thanks{Corresponding author: linhn@cqu.edu.cn}\\
$^{1}$Center for Gravitation and Cosmology, College of Physical Science and Technology, Yangzhou University, Yangzhou 225009, China\\
$^{2}$Department of Physics, Chongqing University, Chongqing 401331, China\\
}

\date{Accepted XXX. Received YYY; in original form ZZZ}

\pubyear{2024}

\begin{document}
\label{firstpage}
\pagerange{\pageref{firstpage}--\pageref{lastpage}}
\maketitle

\begin{abstract}
Fast Radio Bursts (FRBs) are enigmatic radio pulses whose origins are poorly understood. The dispersion measure of host galaxy (${\rm DM_{host}}$) provides critical insights into the local environment of FRB sources. In this study, we analyze a sample of 117 well-localized FRBs to investigate the statistical properties of ${\rm DM_{host}}$ and its potential correlations with host galaxy parameters, including redshift, stellar mass, star formation rate (SFR), and galaxy age. Our results reveal that ${\rm DM_{host}}$ is consistent with a log-normal distribution, with the mean $\mu_{\rm host}=5.03\pm 0.02$ and standard deviation $\sigma_{\rm host} = 0.96\pm 0.03$, which corresponds to a median value ${\rm Med}({\rm DM_{host}})=\exp(\mu_{\rm host})=153\pm 3~{\rm pc~cm^{-3}}$. We find a moderate positive correlation between ${\rm DM_{host}}$ and redshift, but no statistically significant correlations are found between ${\rm DM_{host}}$ and stellar mass, SFR, or galaxy age. Our findings highlight the importance of ${\rm DM_{host}}$ as a diagnostic tool for unraveling FRB origins, and underscore the need for future FRB surveys with deep multiwavelength host galaxy follow-up.
\end{abstract}




\section{Introduction}\label{sec:introduction}

Fast radio bursts (FRBs) are millisecond-duration and highly energetic radio transients of unknown origin, see e.g. \citep{Platts:2018hiy,Petroff:2019tty,Xiao:2021omr,Zhang_2023} for review. Triggering a paradigm shift in radio astronomy since their serendipitous discovery in Parkes telescope archival data \citep{Lorimer:2007qn}, the FRB population has undergone rapid growth in recent years, with over 1000 events have been cataloged\footnote{https://www.wis-tns.org/}. The large dispersion measures (DMs), which exceed the Galactic contribution by orders of magnitude, provide irrefutable evidence for their extragalactic origin. The extragalactic nature was conclusively validated through host galaxy identifications and spectroscopic redshift measurements \citep{Keane:2016yyk,Chatterjee:2017dqg}\footnote{The host galaxy of the first localized FRB 20150418 \citep{Keane:2016yyk} is hightly debated, see e.g. \citep{Williams:2016zys,Li:2016vzg}.}. Observationally, FRBs can be classified in two categories: repeaters and non-repeaters, according to the number of detected bursts from a specific source. This dichotomy suggests diverse possible progenitor mechanisms such as magnetar flares, compact object mergers, or exotic physics \citep{Platts:2018hiy}. Despite rapid observational progress, the origins of FRBs and the astrophysical environments responsible for their emission remain subjects of intense debate.

FRBs serve as exceptional probes for both cosmological parameter constraints and fundamental physics investigations \citep{Bhandari:2021thi,Zhao:2022yiv,Zhang:2024bar,Wu:2024iyu,Glowacki:2024cgu,Zhang:2025thh}. The DM of FRBs, accumulated through interactions with free electrons during intergalactic propagation, provides a direct tracer of the cosmic baryon density. This enables critical tests of structure formation by quantifying the warm-hot intergalactic medium, hence resolves the long-lasting ``missing baryon" problem \citep{McQuinn:2013tmc,Macquart:2020lln,Yang:2022ftm,Connor:2024mjg}. In addition, cross-calibrating DMs with spectroscopically confirmed host galaxy redshifts further establishes an independent cosmic ladder, yielding competitive constraints on the Hubble constant \citep{Li:2017mek,Wu:2021jyk,Hagstotz:2021jzu,James:2022dcx,Gao:2025fcr}. Moreover, FRBs are also useful in tracing the reionization history \citep{Bhattacharya:2020rtf,Pagano:2021zla}, testing the cosmological principles \citep{Qiang:2019zrs,Lin:2021syj}, probing compact dark matter \citep{Munoz:2016tmg}. In the realm of fundamental physics, FRBs provide a unique laboratory for testing foundational theories, such as testing Einstein's equivalence principle \citep{Wei:2015hwd,Tingay:2016tgf,Wu:2017yjl} and constraining photon mass \citep{Wu:2016brq,Shao:2017tuu,Lin:2023jaq}.

The applications of FRBs in cosmology and fundamental physics mainly depend on the DM-redshift relation. However, current cosmological analyses remain fundamentally limited by the inherent degeneracy between different components of DMs. Especially, the poor knowledge on DM of the host galaxy (${\rm DM_{host}}$) introduces a large uncertainty in FRB cosmology. Conventional approaches often rely on oversimplified assumptions, such as assigning fixed host contributions or employing prior distributions when extracting ${\rm DM_{host}}$. However, observations show that the value of ${\rm DM_{host}}$ can vary significantly from burst to burst, ranging from several tens to several hundreds ${\rm pc~cm^{-3}}$ \citep{Tendulkar:2017vuq,Bannister:2019iju,Xu:2021qdn}, or in some extreme cases up to 1000 ${\rm pc~cm^{-3}}$ \citep{Niu_2022,KochOcker:2022ook}. Both numerical simulations and theoretical considerations demonstrate that the probability distribution of ${\rm DM_{host}}$ can be described by the log-normal function \citep{Macquart:2020lln,Zhang:2020mgq}. Moreover, ${\rm DM_{host}}$ encodes important information about the local environment of FRB: It reflects the density of ionized gas in the interstellar medium (ISM) of the host galaxy, the circumgalactic medium (CGM) and near the progenitor itself. Understanding ${\rm DM_{host}}$ is thus essential not only for refining FRB-based cosmology, but also for discriminating between progenitor models and propagation effects.

Previous studies on ${\rm DM_{host}}$ have been limited by small sample sizes and localization uncertainties \citep{Lin:2022afm}. The advent of dedicated FRB surveys with arcsecond-level localization capabilities, such as CHIME/FRB \citep{CHIMEFRB:2021srp}, ASKAP \citep{Shannon:2024pbu} and DSA-110 \citep{Law:2023ibd}), has revolutionized this field, enabling robust host galaxy associations for more than 100 FRBs. This large sample allows, for the first time, systematic investigation of ${\rm DM_{host}}$ distributions and their correlations with host galaxy properties. In this work, we present a comprehensive Bayesian framework to statistically characterize ${\rm DM_{host}}$ using a sample of 117 well-localized FRBs. Our approach explicitly accounts for measurement uncertainties while reconstructing the underlying ${\rm DM_{host}}$ probability distribution. We further investigate correlations between ${\rm DM_{host}}$ and four key  parameters of host galaxy: redshift, stellar mass, star formation rate (SFR), and galaxy age. These parameters are selected based on their potential connections to FRB progenitor environments: redshift traces cosmic evolution, stellar mass and SFR reflect galaxy assembly history, while galaxy age represents the evolutionary stage of the host galaxy.

The remainder of this paper is organized as follows: Section \ref{sec:method} details our Bayesian methodology, including the hierarchical model construction and parameter estimation techniques. Section \ref{sec:results} presents the main results, including the reconstructed ${\rm DM_{host}}$ distribution and its correlations with host galaxy properties. Finally, discussion and conclusions are presented in Section \ref{sec:conclusions}.

\section{Methodology}\label{sec:method}

The propagation of electromagnetic waves in cold plasma induces modifications to the dispersion relation, introducing frequency-dependent light speed. This dispersive phenomenon becomes particularly evident in low-frequency regimes, as exemplified by radio emissions associated with fast radio bursts (FRBs). The resultant frequency-dependent time delay for extragalactic FRB signals traversing the intergalactic medium to Earth can be quantified by the dispersion measure (DM), which is defined by the integral of free electron density along the light path. FRBs at further distance are expected to confront more electrons thus have larger DMs, so it is usually regarded as a proxy of distance. The DM value of a FRB can be measured from the dynamic spectra with negligible uncertainty.

Generally, the observed DM of an extragalactic FRB can be decomposed into four components \citep{Deng:2013aga,Gao:2014iva,Macquart:2020lln},
\begin{equation}\label{eq:DM_obs}
    {\rm DM_{obs}}={\rm DM_{ISM}}+{\rm DM_{halo}}+{\rm DM_{IGM}}+\frac{{\rm DM_{host}}}{1+z}.
\end{equation}
where the first to fourth terms on the right-hand side represent the contribution of Milky Way interstellar medium (ISM), the Milky Way halo, the intergalactic medium between host galaxy and Milky Way, and the FRB host galaxy, respectively. The factor $1+z$ in the last term accounts for the cosmic expansion. Some authors have proposed an additional fifth term ${\rm DM_{source}}$ to account for the contribution from plasma in the FRB's local environment \citep{Zhang:2020mgq}. However, due to current observational limitations, it is difficult to break the degeneracy between ${\rm DM_{host}}$ and ${\rm DM_{source}}$. Therefore, we adopt a conservative approach by absorbing any potential source-related contribution into the ${\rm DM_{host}}$ term for this analysis.

The ${\rm DM_{ISM}}$ term can be estimated using Milky Way electron density models, such as the NE2001 model \citep{Cordes:2002wz} and YMW16 model \citep{Yao_2017msh}. While both models demonstrate consistency in ${\rm DM_{ISM}}$ predictions for FRBs at high Galactic latitudes $(|b|>10^{\circ})$, systematic discrepancies emerge at low Galactic latitudes, where the YMW16 model tends to overestimate ionized gas contributions \citep{KochOcker:2021fia}. So we adopt the NE2001 model as our baseline. It should be noted that the NE2001 model does not provide uncertainty estimates; we therefore conservatively assign 50\% uncertainty on it. In summary, we assume that the ${\rm DM_{ISM}}$ value follows Gaussian distribution, with mean value $\mu_{\rm ISM}={\rm NE2001}$, and standard deviation $\sigma_{\rm ISM}=0.5\times\mu_{\rm ISM}$. The probability distribution of ${\rm DM_{ISM}}$ can be written as 
\begin{equation}
    p_{\rm ISM}({\rm DM_{ISM}})=\frac{1}{\sqrt{2\pi}\sigma_{\rm ISM}}\exp{\left[-\frac{({\rm DM_{ISM}}-\mu_{\rm ISM})^2}{2\sigma_{\rm ISM}^2}\right]}.
\end{equation}

The $\mathrm{DM_{halo}}$ term corresponds to the dispersion measure contribution from ionized gas in the Milky Way halo. This component remains poorly constrained, with widely divergent estimates across the literature. For instance, \citet{Connor:2024mjg} adopted a fixed value of ${\rm DM_{halo} = 30~{\rm pc~cm^{-3}}}$, whereas \citet{Macquart:2020lln} assumed ${\rm DM_{halo} = 50~{\rm pc~cm^{-3}}}$. In contrast, the DSA collaboration used a lower value of ${\rm DM_{halo} = 10~{\rm pc~cm^{-3}}}$ for FRB 20220912A \citep{DeepSynopticArrayTeam:2022rbq}, and a range of ${\rm DM_{halo} = 28\sim 48~{\rm pc~cm^{-3}}}$ for FRB 20220319D \citep{Ravi:2023zfl}. These disparities highlight the significant uncertainty in the Galactic halo’s contribution to the total DM. To avoid inferring unphysical (negative) ${\rm DM_{host}}$ values for some FRBs, adopting a constant $\rm DM_{halo}$ is inappropriate. Following the approach of \cite{Prochaska:2019mkd}, who assumed $\rm DM_{halo} = 50\sim 80~{\rm pc~cm^{-3}}$, we employ a Gaussian prior centered at $\mu_{\rm halo}=65~{\rm pc~cm^{-3}}$ with a standard deviation of $\sigma_{\rm halo}=15~{\rm pc~cm^{-3}}$, consistent with that used by \citet{Yang:2022ftm}. Thus, the distribution of $\rm DM_{halo}$ can be expressed as
\begin{equation}
    p_{\rm halo}({\rm DM_{halo}})=\frac{1}{\sqrt{2\pi}\sigma_{\rm halo}}\exp{\left[-\frac{({\rm DM_{halo}}-\mu_{\rm halo})^2}{2\sigma_{\rm halo}^2}\right]}.
\end{equation}

The ${\rm DM_{IGM}}$ term can be calculated given a specific cosmological model. In the standard ${\rm\Lambda}$CDM model, the average value of ${\rm DM_{IGM}}$ can be written as \citep{Deng:2013aga}
\begin{equation}\label{eq:DM_IGM}
    \langle{\rm DM_{IGM}}(z)\rangle=\frac{3cH_0\Omega_bf_{\rm IGM}f_{e}}{8\pi Gm_p}\int_0^z\frac{1+z}{\sqrt{\Omega_m(1+z)^3+\Omega_\Lambda}}dz,
\end{equation}
where $c$ is the speed of light, $m_p$ is the proton mass, $G$ is the Newtonian gravitational constant, $f_{\rm IGM}$ is the baryon mass fraction in IGM, and $f_e$ is the electron fraction. Several independent baryon census yields consistent result of $f_{\rm IGM}\approx 0.83$ \citep{Fukugita:1997bi,Shull:2011aa}. Adopting the standard Big Bang nucleosynthesis framework where baryon matter consists of $X=3/4$ hydrogen mass fraction and $Y=1/4$ helium mass fraction, and assuming complete ionization of both elements in the low-redshift IGM, we have $f_e=7/8$. The other cosmological parameters are fixed to the Planck 2018 results, i.e. $H_0=67.4~{\rm km~s^{-1}~Mpc^{-1}}$, $\Omega_m=0.315$, $\Omega_\Lambda=0.685$ and $\Omega_{b}=0.0493$ \citep{Aghanim:2018eyx}.

Note that the actual value of ${\rm DM_{IGM}}$ may significantly deviate from equation (\ref{eq:DM_IGM}) due to the fluctuation of matter density. Numerical simulations demonstrate that the distribution of $\rm DM_{IGM}$ can be well modeled by the quasi-Gaussian function \citep{Macquart:2020lln,Zhang:2020xoc},
\begin{equation}\label{eq:P_IGM}
	p_{\rm IGM}(\Delta)=A\Delta^{-\beta}\exp\left[-\frac{(\Delta^{-\alpha}-C_0)^2}{2\alpha^2\sigma_{\rm IGM}^2}\right],
\end{equation}
where $\rm \Delta \equiv DM_{IGM}/\left \langle DM_{IGM} \right \rangle$, $\sigma_{\rm IGM}=Fz^{-1/2}$ is the effective standard deviation with $F\approx 0.2$, $A$ is the normalization constant, and $C_0$ is chosen to ensure that the mean value of $\Delta$ is unity. \citet{Macquart:2020lln} showed that $\alpha = \beta = 3$ yields a reasonable fit, so we also fix these two parameters.

The last term $\rm DM_{host}$ is the most poorly known term, primarily stemming to the lack of comprehensive observations of the local surroundings of FRB sources. However, given the distributions of the former three terms, the distribution of ${\rm DM_{host}}$ can be derived through the convolution,
\begin{eqnarray}\label{eq:P_host}
   p_{\rm host}({\rm DM_{host}})=\iint_0^{\rm DM_{obs}} p_{\rm ISM}({\rm DM_{ISM}})p_{\rm halo}({\rm DM_{halo}}) \nonumber\\
  \times p_{\rm IGM}({\rm DM_{obs}}-{\rm DM_{ISM}}-{\rm DM_{halo}}-\frac{\rm DM_{host}}{1+z})d{\rm DM_{ISM}}d{\rm DM_{halo}}.
\end{eqnarray}
From physical considerations, all the four componential DMs are restricted to be positive definite.

\section{Data and results}\label{sec:results}

We individually calculate the probability distributions of ${\rm DM_{host}}$ for 117 well-localized FRBs using Equation (\ref{eq:P_host}), with the corresponding median values and $1\sigma$ uncertainties summarized in the penultimate column of Table \ref{tab:localized_FRBs} in the Appendix. Figure \ref{fig:DM_distribution} presents some representative cases: FRB 20121102A, FRB 20190520B, and FRB 20200120E. For each source, we systematically display the probability distributions of four distinct DM components, alongside the total observed DM for comparative analysis. All probability densities are normalized such that the area under each probability curve is unity. The vertical red-dashed lines represent the 16th, 50th, and 84th percentiles of the probability distribution of ${\rm DM_{host}}$, respectively. The vertical solid-black line denotes the observed DM value. Note that the intrinsic ${\rm DM_{host}}$ may exceed ${\rm DM_{obs}}$ (as is seen in FRB 20190520B), because the former is scaled by a factor of $1/(1+z)$ in the observer frame.

\begin{figure*}
	\centering
	\includegraphics[width=0.48\textwidth]{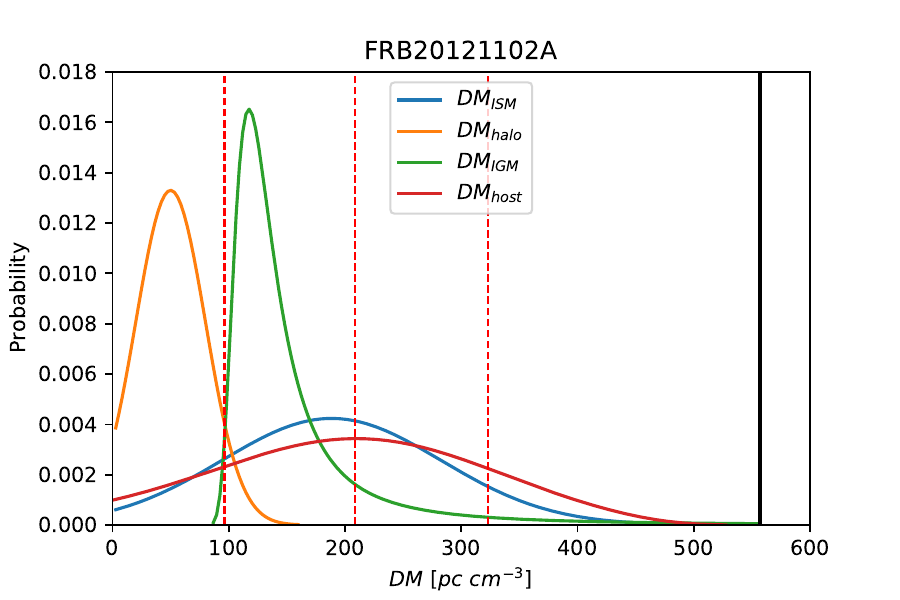}
	\includegraphics[width=0.48\textwidth]{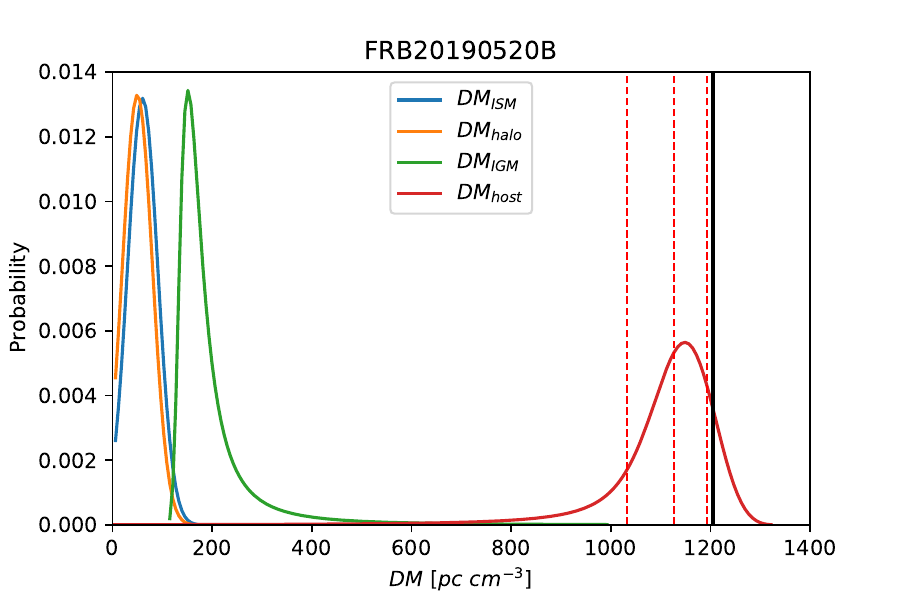}	
    \includegraphics[width=0.48\textwidth]{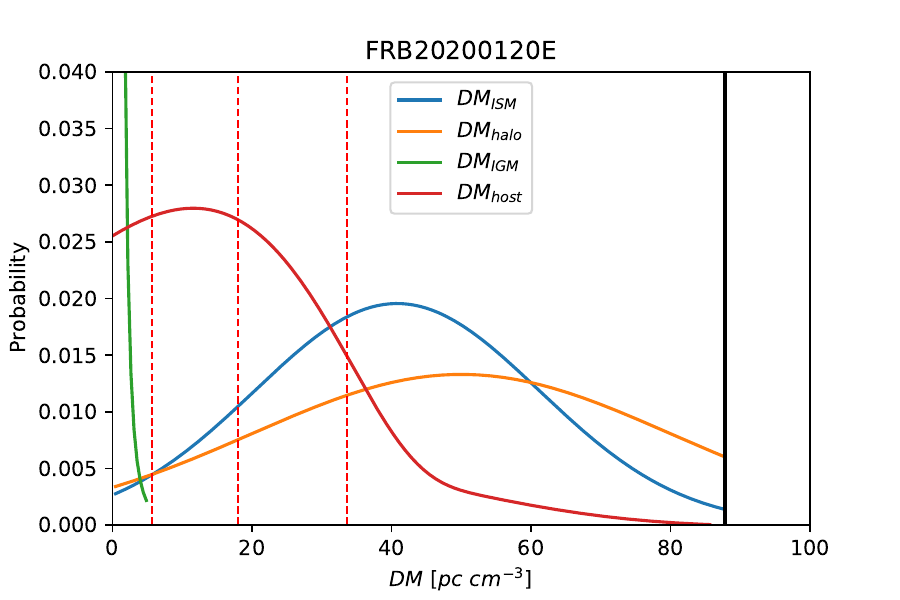}
    \includegraphics[width=0.48\textwidth]{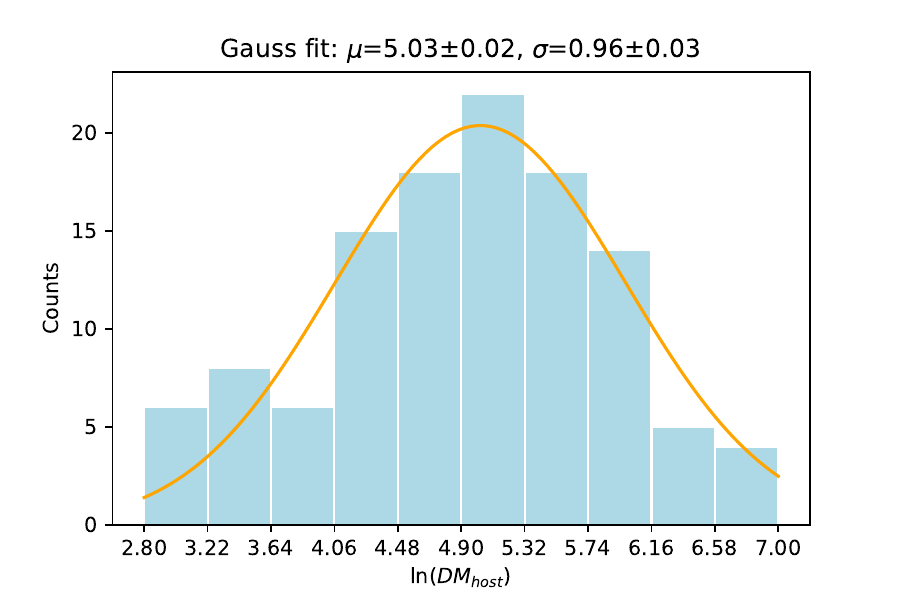}
    \caption{The probability distribution of each component of DM for FRB 20121102A (upper left), FRB 20190520B (upper right) and FRB 20200120E (lower-left). The vertical red-dashed lines are the 16th, 50th and 84th percentiles of the probability distribution of ${\rm DM_{host}}$, respectively. The vertical black-solid line is the value of ${\rm DM_{obs}}$, which has a negligible uncertainty. Lower-right: Histogram of $\ln{\rm DM_{host}}$ with the best-fitting Gaussian distribution superimposed.}\label{fig:DM_distribution}
\end{figure*}

FRB 20121102A is the first discovered repeating FRB locating at redshift $z=0.193$, with the observed DM value ${\rm DM_{obs}=557~pc~cm^{-3}}$ \citep{Tendulkar:2017vuq}. Its host galaxy has been localized to subarcsecond precision in the J2000 coordinates ${\rm (RA,DEC)}=(05^{\rm h}31^{\rm m}58.70^{\rm s}, + 33^{\circ}08'52.5'')$ \citep{Chatterjee:2017dqg}. The NE2001 model predicts a Galactic ISM contribution of ${\rm DM_{ISM}=188~pc~cm^{-3}}$. According to equation (\ref{eq:P_host}), we obtain ${\rm DM_{host} = 208.66_{-112.34}^{+114.14}~pc~cm^{-3}}$, which falls within the typical range for FRB host galaxies. The upper-left panel of Figure \ref{fig:DM_distribution} comparatively displays the probability distributions of the four distinct DM components for this source.

FRB 20190520B is a repeating fast radio burst associated with a persistent radio source at redshift $z=0.241$ \citep{Niu_2022}. This FRB exhibits an exceptionally high DM value (${\rm DM_{obs}=1204.7~pc~cm^{-3}}$) for its redshift, implying that the host galaxy contributes a large amount of DM. The probability distributions of four DM components are shown in the upper-right panel of Figure \ref{fig:DM_distribution}. The NE2001 model predicts ${\rm DM_{ISM}=60.5~pc~cm^{-3}}$. Our Bayesian analysis via equation (\ref{eq:P_host}) yields a host galaxy contribution ${\rm DM_{host} = 1127.21_{-94.72}^{+65.16} ~pc~cm^{-3}}$, which corresponds to ${\rm DM_{host}/(1+z) = 908.31_{-76.33}^{+52.51} ~pc~cm^{-3}}$ in the observer frame, aligns with an independent estimation $903_{-111}^{+72} {\rm~pc~cm^{-3}}$ given by \citet{Niu_2022}. The extremely large value of ${\rm DM_{host}}$ suggests an exceptionally dense local environment of the progenitor.

FRB 20200120E is a repeating fast radio burst hosted by a globular cluster associated with the nearby galaxy M81, which lies at a distance of 3.6 Mpc \citep{Kirsten:2021llv}. The probability distributions of four DM components for this burst are shown in the lower-left panel of Figure \ref{fig:DM_distribution}. FRB 20200120E exhibits several remarkable characteristics: (1) it has the lowest known redshift ($z=0.0008$); (2) its host galaxy stellar mass $(\log(M_*/M_\odot) = 5.77_{-0.22}^{+0.19})$ is at least two orders of magnitude smaller than those of others; and (3) it exhibits the lowest ${\rm DM_{host}}$ value. \citet{Kirsten:2021llv} estimated the host DM to be approximately $15 - 50~{\rm pc~cm^{-3}}$, which closely aligns with our derived values (${\rm DM_{host}}=17.98_{-12.34}^{+15.68}~{\rm pc~cm^{-3}}$). The relatively low ${\rm DM_{host}}$ for this source suggests a low-density local environment.

The inferred ${\rm DM_{host}}$ values for the 117 well-localized FRBs show a spread of three orders of magnitude, ranging from $17.98_{-12.34}^{+15.68}~{\rm pc~cm^{-3}}$ for FRB 20200120E to $1127.21_{-94.72}^{+65.16}{\rm ~pc~cm^{-3}}$ for FRB 20190520B. In addition to FRB 20190520B, two other sources, FRB 20220610A and FRB 20240123A, exhibit rest-frame ${\rm DM_{host}}$ exceeding $1000 {\rm ~pc~cm^{-3}}$, suggesting a highly complex (e.g., dense or turbulent) local environment for these systems. To visualize the probability distribution of ${\rm DM_{host}}$, we divide the (natural based) logarithmic values $\ln {\rm DM_{host}}$ (the 50th percentile values) into 10 uniform bins and plot the histogram in the lower-right panel of Figure \ref{fig:DM_distribution}. The histogram can be well fitted by the normal distribution
\begin{equation}
  p(\ln\rm DM_{ host})=\frac{1}{\sqrt{2\pi}\sigma_{\rm host}}\exp\left[-\frac{(\ln {\rm DM_{host}}-\mu_{\rm host})^2}{2\sigma_{\rm host}^2}\right],
\end{equation}
with the best-fitting parameters $\sigma_{\rm host}=0.96\pm 0.03$ and $\mu=5.03\pm 0.02$. This corresponds to a median value ${\rm Med}({\rm DM_{host}})=\exp(\mu_{\rm host})=153\pm 3~{\rm pc~cm^{-3}}$. Our finding provide independent observational validation of cosmological simulations predicting that ${\rm DM_{host}}$ follows the log-normal distribution \citep{Macquart:2020lln,Zhang:2020mgq}.

To assess the goodness-of-fit of the log-normal model, we employ the Kolmogorov-Smirnov (K-S) test via the Python module {\it scipy.stats.kstest}. The test compares the distribution of the inferred $\ln(\mathrm{DM_{host}})$ values against a theoretical normal distribution parameterized by the fitted parameters. The resultant high p-value ($p = 0.92$) suggests no evidence against the null hypothesis of normality. To further evaluate the robustness of this conclusion against potential outliers, we perform a bootstrap analysis with 10000 iterations. In each iteration, we generate a new sample of the same size as the original dataset by resampling the original $\ln(\mathrm{DM_{host}})$ values with replacement, and compute the K-S test p-value against a normal distribution parameterized by the sample's mean and standard deviation. The resulting 95\% confidence interval for the p-value is [0.09, 0.95], which lies entirely above the conventional significance level of 0.05, confirming that the result is robust.

\begin{figure*}
	\centering
	\includegraphics[width=0.48\textwidth]{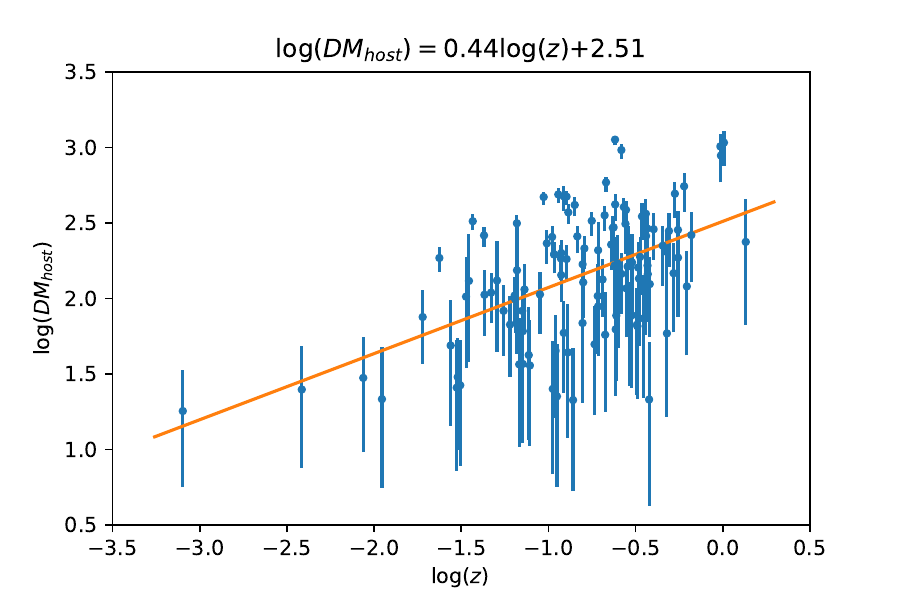}
	\includegraphics[width=0.48\textwidth]{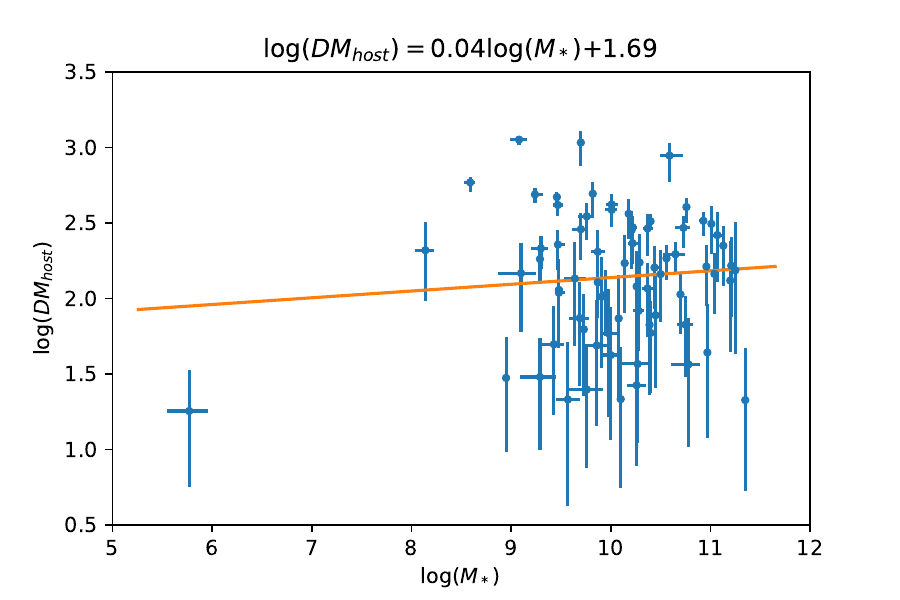}
    \includegraphics[width=0.48\textwidth]{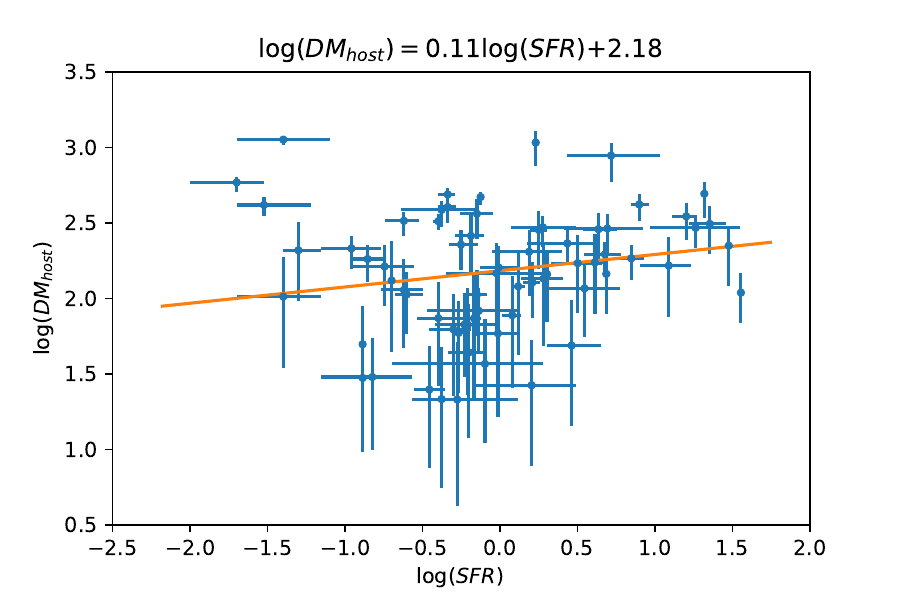}	
    \includegraphics[width=0.48\textwidth]{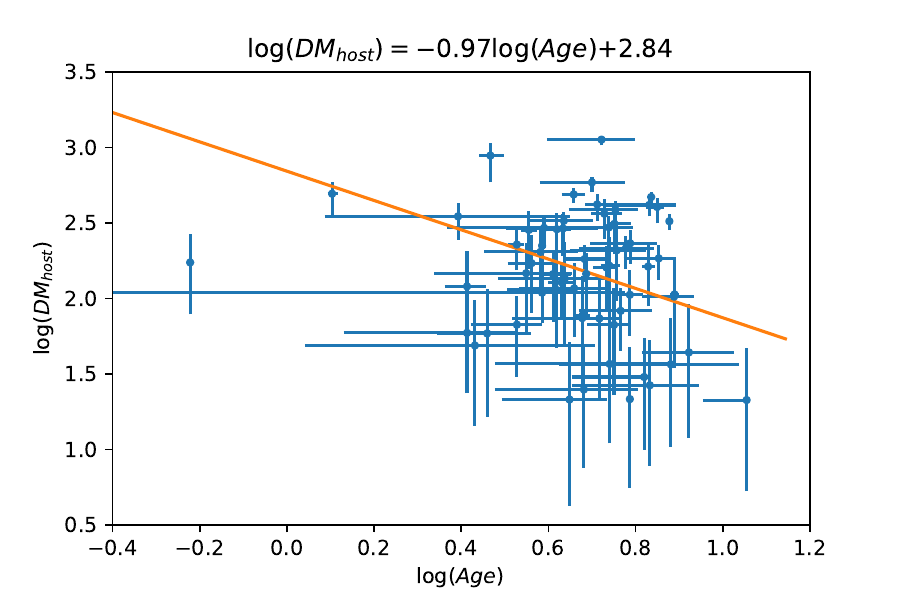}	
    \caption{The correlations between ${\rm DM_{host}}$ and the properties of host galaxies (redshift, stellar mass, SFR and galaxy age). The errorbar represents the $1\sigma$ uncertainty. The orange line represents the linear regression line calculated via the orthogonal distance regression method.}\label{fig:correlations}
\end{figure*}

To investigate the potential correlations between ${\rm DM_{host}}$ and the fundamental properties of host galaxies (redshift, stellar mass, SFR and galaxy age), we systematically examine the following four bivariate correlations,
\begin{eqnarray}
  \log {\rm DM_{host}} &=& k \log z + m,\\
  \log {\rm DM_{host}} &=& k \log M_* + m,\\
  \log {\rm DM_{host}} &=& k \log {\rm SFR} + m,\\
  \log {\rm DM_{host}} &=& k \log {\rm Age} + m.
\end{eqnarray}
These correlations are visualized in the scatter plots of Figure \ref{fig:correlations}, along with corresponding orthogonal distance regression (ODR) fits. The ODR method is employed to appropriately account for measurement uncertainties in both the $x$ and $y$ directions during regression process. For data points with asymmetric uncertainties, we symmetrize them using the root mean square. Data points without reported uncertainties are assigned the average uncertainty value across the dataset. We note that the ${\rm DM_{host}}$ values span three orders of magnitude and that their uncertainties are not significantly correlated with ${\rm DM_{host}}$ itself. However, when propagated to the logarithmic scale, the data points with larger ${\rm DM_{host}}$ values exhibit smaller uncertainties. As a result, the ODR fitting is dominated by points with high ${\rm DM_{host}}$ values. To avoid this issue, we assign the same uncertainty of $\ln(\rm DM_{host})$ (the mean value) to each data point during the fitting process. The ODR fit is performed using the Python module \textit{scipy.odr}, and the best-fitting parameters are summarized in Table \ref{tab:parameters}.

\begin{table*}
    \centering
    \caption{\small{The best-fitting parameters, together with the Pearson's correlation coefficients and p-values.}}\label{tab:parameters}
    \begin{tabular}{llll}
    \hline
    correlation & parameters & Pearson & p-value \\
    \hline
    $\log {\rm DM_{host}}=k \log z + m$              & $k=+0.44\pm 0.07, m=2.51\pm 0.06$ & $r=+0.52$ & $1.73\times 10^{-9}$\\
    $\log {\rm DM_{host}}=k \log M_* + m$            & $k=+0.04\pm 0.06, m=1.69\pm 0.61$ & $r=+0.08$ & 0.48\\
    $\log {\rm DM_{host}}=k \log {\rm SFR} + m$      & $k=+0.11\pm 0.07, m=2.18\pm 0.05$ & $r=+0.14$ & 0.24\\
    $\log {\rm DM_{host}}=k \log {\rm Age} + m$      & $k=-0.97\pm 0.26, m=2.84\pm 0.18$ & $r=-0.18$ & 0.15\\
    \hline
    \end{tabular}
\end{table*}

To quantify the strength of the correlations and its statistical significance, we compute Pearson’s correlation coefficients and p-values using the Python module {\it scipy.stats.pearsonr}. The results are presented in the last two columns of Table \ref{tab:parameters}. We identify moderate positive correlations between ${\rm DM_{host}}$ and redshift ($r=+0.52$), while finding no significant dependence on stellar mass ($r=+0.08$), SFR ($r=+0.14$) or galactic age ($r=-0.18$). The small p-value for the ${\rm DM_{host}-redshift}$ correlation indicates that it is highly unlikely the observed relationship occurred by chance. For the remaining three correlations, however, the large p-values ($p>0.1$) suggest that there is no strong evidence for a statistically significant relationship. These results challenge conventional models that attribute ${\rm DM_{host}}$ variations primarily to galaxy mass or SFR scaling laws, but instead suggesting that cosmic redshift evolution plays dominant roles in shaping host galaxy contributions.

\begin{figure}
	\centering
	\includegraphics[width=0.48\textwidth]{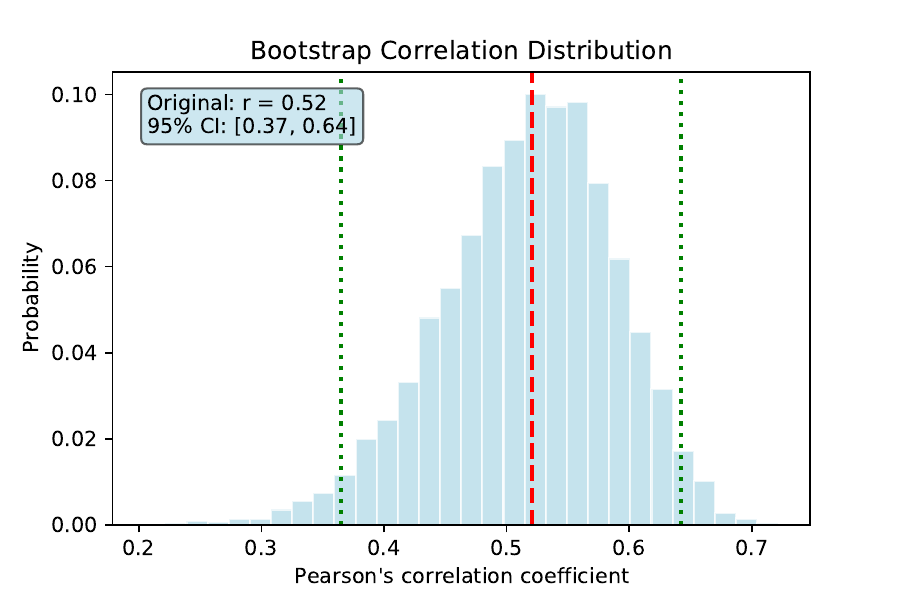}	
    \caption{Distribution of Pearson's correlation coefficients from 10000 bootstrap samples. The 95\% confidence interval is indicated by the dotted green lines, and the correlation coefficient from the original data is marked by the dashed red line.}\label{fig:bootstrap}
\end{figure}

To determine whether the correlation between ${\rm DM_{host}}$ and redshift is influenced by outliers, we conduct a bootstrap resampling analysis. The procedure is as follows: We generate 10000 bootstrap samples by randomly resampling the data with replacement. For each sample, we compute Pearson’s correlation coefficient. The distribution of the correlation coefficients from these 10000 samples is shown in Figure \ref{fig:bootstrap}. The 95\% confidence interval for the Pearson’s correlation coefficient is $[0.37, 0.64]$. This interval encompasses the correlation coefficient obtained from the original data ($r=0.52$) and lies entirely above zero, providing strong evidence that the observed linear correlation is not driven by a small number of influential outliers.

The estimation of ${\rm DM_{host}}$ relies on assumptions regarding the distribution of ${\rm DM_{IGM}}$ and its redshift evolution; hence, redshift is incorporated in the calculation of ${\rm DM_{host}}$. Notably, the distribution of ${\rm DM_{IGM}}$, as expressed in equation (\ref{eq:P_IGM}), is asymmetric around its median at a given redshift. To assess whether the assumed distribution shape influences our results, we adopt an alternative model by assuming ${\rm DM_{IGM}}$ follows a Gaussian distribution. The mean is given by equation (\ref{eq:DM_IGM}), and the standard deviation is set to 20\% of the mean. Using this Gaussian model, we recalculated ${\rm DM_{host}}$ for each FRB and found that the resulting histogram remains consistent with that shown in the lower-right panel of Figure \ref{fig:DM_distribution}. A Gaussian fit yields parameter values of $\mu = 4.80 \pm 0.03$ and $\sigma = 1.06 \pm 0.03$. In this case, a linear correlation between ${\rm DM_{host}}$ and redshift remains evident. The ODR regression yields parameters $k = 0.37 \pm 0.07$ and $m = 2.38 \pm 0.07$, with a Pearson's correlation coefficient of $r = 0.44$ and a p-value of $p = 7.38 \times 10^{-7}$. These results confirm that the correlation between ${\rm DM_{host}}$ and redshift is robust and not strongly influenced by the specific distribution assumed for ${\rm DM_{IGM}}$.

\section{Discussion and Conclusions}\label{sec:conclusions}

Based on a sample of 117 well-localized FRBs, we present a systematic investigation of DMs of FRB host galaxies (${\rm DM_{host}}$) using Bayesian statistical methods. Our analysis reveals that ${\rm DM_{host}}$ is well described by a log-normal distribution, peaking at $\sim 153~{\rm pc~cm^{-3}}$, consistent with that obtained by \citet{Connor:2024mjg}. This result provides independent observational confirmation of the log-normal distribution for host DM, as predicted by cosmological simulations \citep{Macquart:2020lln, Zhang:2020mgq}. The broad log-normal distribution of ${\rm DM_{host}}$ underscores the need to account for host galaxy contributions as a significant uncertainty in FRB-based cosmology. A moderate and statistical significant positive correlation is found between ${\rm DM_{host}}$ and redshift (Pearson's correlation coefficient $r = 0.52$, and p-value $1.73\times 10^{-9}$), suggesting that high-redshift galaxies tend to have systematically larger dispersion measures. However, no statistically significant correlations are found between ${\rm DM_{host}}$ and stellar mass, SFR, or galaxy age. The lack of strong correlations with host galaxy properties complicates the precise modeling of ${\rm DM_{host}}$.

Our findings suggest that the ${\rm DM_{host}}$ variations are not predominantly governed by the global properties of the host galaxy. Instead, the significant correlation with redshift points towards evolutionary effects, potentially linked to the changing metallicity or density of the circum-burst environment over cosmic time. The lack of correlation with stellar mass, SFR, and age implies that the local environment immediately surrounding the FRB progenitor, such as a dense stellar cluster, supernova remnant, or the progenitor system's own ejecta, may play a more decisive role in contributing to ${\rm DM_{host}}$ than the integrated properties of the host galaxy ISM. This underscores the importance of progenitor models and local environmental conditions in determining the observed dispersion measure, favoring scenarios where the local FRB environment provides a significant and potentially dominant contribution to ${\rm DM_{host}}$.

Previous studies have investigated correlations between $\rm DM_{host}$ and host galaxy properties from both theoretical and observational perspectives. Using Illustris and IllustrisTNG cosmological simulations, \citet{Mo:2022qxz} found a positive correlation between $\rm DM_{host}$ and host galaxy stellar mass or redshift. Based on 12 well-localized FRBs, \citet{Bernales_Cortes_2025} employed two different methods to estimate $\rm DM_{host}$, one directly using VLT/MUSE observations of the host galaxies, and the other applying an indirect approach based on the Macquart relation. Their results from both methods were consistent, indicating significant positive correlations between $\rm DM_{host}$ and both stellar mass and SFR, but no strong correlation with redshift. In contrast, our analysis shows that $\rm DM_{host}$ is moderately correlated with redshift, but exhibits no significant correlation with stellar mass, SFR, or galaxy age. These findings appear inconsistent with each other. The discrepancies may stem from several factors. First, \citet{Mo:2022qxz} relied on cosmological simulations, which are sensitive to the adopted parameter settings, whereas \citet{Bernales_Cortes_2025} and our work are based on observational FRB samples, making direct comparisons with simulation-based results difficult. Second, the sample size differs substantially: \citet{Bernales_Cortes_2025} used 12 FRBs, while our study includes 117 FRBs. An order-of-magnitude increase in sample size could significantly affect the inferred correlations. Third, when applying the Macquart relation, \citet{Bernales_Cortes_2025} used average values for each DM component ($\rm DM_{ISM}$, $\rm DM_{halo}$, $\rm DM_{IGM}$) without fully accounting for their probability distributions, which may introduce bias. In summary, methodological differences and variations in data samples likely contribute to the divergent results. Future studies with larger, uniformly processed FRB samples may help resolve these discrepancies and lead to more conclusive insights.

The FRB sample utilized in this study comprises data collected from multiple observational facilities. As a result, the host galaxy properties were derived using heterogeneous methodologies, which may introduce systematic biases. For instance, star formation rates (SFRs) are commonly estimated by combining multi-band photometric and spectroscopic data with stellar population synthesis (SPS) models to perform spectral energy distribution (SED) fitting, thereby reconstructing non-parametric star formation histories \citep{Gordon:2023cgw}. Alternatively, SFR can be derived directly from H$\alpha$ emission line luminosity measurements \citep{Heintz2020}. In this work, since we focus on the cosmological ensemble properties of FRB hosts by synthesizing previously published measurements, we adopt the values as reported in the original reference papers without further homogenization. Although a detailed analysis of the systematic differences among various derivation techniques is undoubtedly important, such an investigation lies beyond the scope of the current study. We argue that, despite these potential systematic uncertainties, the large-scale and statistically significant trends identified in our analysis remain robust.

The significant correlation between $\rm DM_{host}$ and redshift has important implications for FRB population models. This correlation could stem from an intrinsic evolution of host galaxy environments with cosmic time, or it could arise from selection effects inherent to current FRB surveys. A rigorous distinction between these scenarios requires comprehensive population synthesis models that incorporate survey selection functions, which is beyond our present work. Future efforts should prioritize expanding FRB samples, particularly at $z>1$, and acquiring multi-wavelength data (e.g., ionized gas kinematics, X-ray emission) to elucidate the physical drivers of ${\rm DM_{host}}$ variations. Advanced techniques for decomposing ${\rm DM_{host}}$, such as incorporating host galaxy morphology and inclination corrections, could break degeneracies between local FRB environments and global galactic properties. Joint analyses of ${\rm DM_{host}}$ statistics and FRB polarization properties may further illuminate the role of magnetized media in burst production. Upcoming facilities such as CHIME, ASKAP, and SKA will enable systematic studies of FRB host galaxies across cosmic epochs, offering unprecedented opportunities to unify progenitor models and probe the baryon cycle in galaxies.

\section*{Acknowledgements}
This work has been supported by the National Natural Science Fund of China under grant nos. 12275034, 12347101 and 12175192, and the Natural Science Fund of Chongqing under grant no. CSTB2022NSCQ-MSX0357.

\section*{Data Availability}
The data underlying this article will be shared on reasonable request to the corresponding author.



\bibliographystyle{mnras}
\bibliography{reference} 

\begin{thebibliography}{}
\makeatletter
\relax
\def\mn@urlcharsother{\let\do\@makeother \do\$\do\&\do\#\do\^\do\_\do\%\do\~}
\def\mn@doi{\begingroup\mn@urlcharsother \@ifnextchar [ {\mn@doi@} {\mn@doi@[]}}
\def\mn@doi@[#1]#2{\def\@tempa{#1}\ifx\@tempa\@empty \href {http://dx.doi.org/#2} {doi:#2}\else \href {http://dx.doi.org/#2} {#1}\fi \endgroup}
\def\mn@eprint#1#2{\mn@eprint@#1:#2::\@nil}
\def\mn@eprint@arXiv#1{\href {http://arxiv.org/abs/#1} {{\tt arXiv:#1}}}
\def\mn@eprint@dblp#1{\href {http://dblp.uni-trier.de/rec/bibtex/#1.xml} {dblp:#1}}
\def\mn@eprint@#1:#2:#3:#4\@nil{\def\@tempa {#1}\def\@tempb {#2}\def\@tempc {#3}\ifx \@tempc \@empty \let \@tempc \@tempb \let \@tempb \@tempa \fi \ifx \@tempb \@empty \def\@tempb {arXiv}\fi \@ifundefined {mn@eprint@\@tempb}{\@tempb:\@tempc}{\expandafter \expandafter \csname mn@eprint@\@tempb\endcsname \expandafter{\@tempc}}}

\bibitem[\protect\citeauthoryear{Aghanim et~al.}{Aghanim et~al.}{2020}]{Aghanim:2018eyx}
Aghanim N.,  et~al., 2020, \mn@doi [Astron. Astrophys.] {10.1051/0004-6361/201833910}, 641, A6

\bibitem[\protect\citeauthoryear{Amiri et~al.}{Amiri et~al.}{2021}]{CHIMEFRB:2021srp}
Amiri M.,  et~al., 2021, \mn@doi [Astrophys. J. Supp.] {10.3847/1538-4365/ac33ab}, 257, 59

\bibitem[\protect\citeauthoryear{Amiri et~al.}{Amiri et~al.}{2025}]{Amiri:2025sbi}
Amiri M.,  et~al., 2025, \mn@doi [Astrophys. J. Suppl.] {10.3847/1538-4365/addbda}, 280, 6

\bibitem[\protect\citeauthoryear{Andersen et~al.}{Andersen et~al.}{2023}]{CHIMEFRB:2023myn}
Andersen B.~C.,  et~al., 2023, \mn@doi [Astrophys. J.] {10.3847/1538-4357/acc6c1}, 947, 83

\bibitem[\protect\citeauthoryear{Bannister et~al.}{Bannister et~al.}{2019}]{Bannister:2019iju}
Bannister K.~W.,  et~al., 2019, \mn@doi [Science] {10.1126/science.aaw5903}, 365, 565

\bibitem[\protect\citeauthoryear{Bernales-Cortes, Tejos, Prochaska, Khrykin, Marnoch, Ryder  \& Shannon}{Bernales-Cortes et~al.}{2025}]{Bernales_Cortes_2025}
Bernales-Cortes L.,  Tejos N.,  Prochaska J.~X.,  Khrykin I.~S.,  Marnoch L.,  Ryder S.~D.,   Shannon R.~M.,  2025, \mn@doi [Astron. Astrophys.] {10.1051/0004-6361/202452026}, 696, A81

\bibitem[\protect\citeauthoryear{Bhandari \& Flynn}{Bhandari \& Flynn}{2021}]{Bhandari:2021thi}
Bhandari S.,  Flynn C.,  2021, \mn@doi [Universe] {10.3390/universe7040085}, 7, 85

\bibitem[\protect\citeauthoryear{Bhandari et~al.}{Bhandari et~al.}{2020}]{Bhandari:2020oyb}
Bhandari S.,  et~al., 2020, \mn@doi [Astrophys. J. Lett.] {10.3847/2041-8213/ab672e}, 895, L37

\bibitem[\protect\citeauthoryear{Bhandari et~al.}{Bhandari et~al.}{2022}]{Bhandari:2021pvj}
Bhandari S.,  et~al., 2022, \mn@doi [Astron. J.] {10.3847/1538-3881/ac3aec}, 163, 69

\bibitem[\protect\citeauthoryear{Bhandari et~al.}{Bhandari et~al.}{2023}]{Bhandari:2022ton}
Bhandari S.,  et~al., 2023, \mn@doi [Astrophys. J.] {10.3847/1538-4357/acc178}, 948, 67

\bibitem[\protect\citeauthoryear{Bhardwaj et~al.}{Bhardwaj et~al.}{2021a}]{Bhardwaj:2021xaa}
Bhardwaj M.,  et~al., 2021a, \mn@doi [Astrophys. J. Lett.] {10.3847/2041-8213/abeaa6}, 910, L18

\bibitem[\protect\citeauthoryear{Bhardwaj et~al.}{Bhardwaj et~al.}{2021b}]{Bhardwaj:2021hgc}
Bhardwaj M.,  et~al., 2021b, \mn@doi [Astrophys. J. Lett.] {10.3847/2041-8213/ac223b}, 919, L24

\bibitem[\protect\citeauthoryear{Bhardwaj et~al.}{Bhardwaj et~al.}{2024}]{Bhardwaj:2023vha}
Bhardwaj M.,  et~al., 2024, \mn@doi [Astrophys. J. Lett.] {10.3847/2041-8213/ad64d1}, 971, L51

\bibitem[\protect\citeauthoryear{Bhattacharya, Kumar  \& Linder}{Bhattacharya et~al.}{2021}]{Bhattacharya:2020rtf}
Bhattacharya M.,  Kumar P.,   Linder E.~V.,  2021, \mn@doi [Phys. Rev. D] {10.1103/PhysRevD.103.103526}, 103, 103526

\bibitem[\protect\citeauthoryear{Caleb et~al.}{Caleb et~al.}{2023}]{Caleb:2023atr}
Caleb M.,  et~al., 2023, \mn@doi [Mon. Not. Roy. Astron. Soc.] {10.1093/mnras/stad1839}, 524, 2064

\bibitem[\protect\citeauthoryear{Cassanelli et~al.}{Cassanelli et~al.}{2024}]{Cassanelli:2023hvg}
Cassanelli T.,  et~al., 2024, \mn@doi [Nature Astron.] {10.1038/s41550-024-02357-x}, 8, 1429

\bibitem[\protect\citeauthoryear{Chatterjee et~al.}{Chatterjee et~al.}{2017}]{Chatterjee:2017dqg}
Chatterjee S.,  et~al., 2017, \mn@doi [Nature] {10.1038/nature20797}, 541, 58

\bibitem[\protect\citeauthoryear{Connor et~al.}{Connor et~al.}{2025}]{Connor:2024mjg}
Connor L.,  et~al., 2025, \mn@doi [Nature Astron.] {10.1038/s41550-025-02566-y}, 9, 1226

\bibitem[\protect\citeauthoryear{Cordes \& Lazio}{Cordes \& Lazio}{2002}]{Cordes:2002wz}
Cordes J.~M.,  Lazio T. J.~W.,  2002, \mn@doi [arXiv: astro-ph/0207156] {10.48550/arXiv.astro-ph/0207156}

\bibitem[\protect\citeauthoryear{Deng \& Zhang}{Deng \& Zhang}{2014}]{Deng:2013aga}
Deng W.,  Zhang B.,  2014, \mn@doi [Astrophys. J. Lett.] {10.1088/2041-8205/783/2/L35}, 783, L35

\bibitem[\protect\citeauthoryear{Driessen et~al.}{Driessen et~al.}{2023}]{Driessen:2023lxj}
Driessen L.~N.,  et~al., 2023, \mn@doi [Mon. Not. Roy. Astron. Soc.] {10.1093/mnras/stad3329}, 527, 3659

\bibitem[\protect\citeauthoryear{Eftekhari et~al.}{Eftekhari et~al.}{2025}]{Eftekhari:2024bmi}
Eftekhari T.,  et~al., 2025, \mn@doi [Astrophys. J. Lett.] {10.3847/2041-8213/ad9de2}, 979, L22

\bibitem[\protect\citeauthoryear{Fong et~al.}{Fong et~al.}{2021}]{Fong:2021xxj}
Fong W.-f.,  et~al., 2021, \mn@doi [Astrophys. J. Lett.] {10.3847/2041-8213/ac242b}, 919, L23

\bibitem[\protect\citeauthoryear{Fukugita, Hogan  \& Peebles}{Fukugita et~al.}{1998}]{Fukugita:1997bi}
Fukugita M.,  Hogan C.~J.,   Peebles P. J.~E.,  1998, \mn@doi [Astrophys. J.] {10.1086/306025}, 503, 518

\bibitem[\protect\citeauthoryear{Gao, Li  \& Zhang}{Gao et~al.}{2014}]{Gao:2014iva}
Gao H.,  Li Z.,   Zhang B.,  2014, \mn@doi [Astrophys. J.] {10.1088/0004-637X/788/2/189}, 788, 189

\bibitem[\protect\citeauthoryear{Gao, Wu, Hu, Yi, Zhou, Wang  \& Dai}{Gao et~al.}{2025}]{Gao:2025fcr}
Gao D.~H.,  Wu Q.,  Hu J.~P.,  Yi S.~X.,  Zhou X.,  Wang F.~Y.,   Dai Z.~G.,  2025, \mn@doi [Astron. Astrophys.] {10.1051/0004-6361/202453006}, 698, A215

\bibitem[\protect\citeauthoryear{Glowacki \& Lee}{Glowacki \& Lee}{2024}]{Glowacki:2024cgu}
Glowacki M.,  Lee K.-G.,  2024, arXiv:2410.24072

\bibitem[\protect\citeauthoryear{Gordon et~al.}{Gordon et~al.}{2023}]{Gordon:2023cgw}
Gordon A.~C.,  et~al., 2023, \mn@doi [Astrophys. J.] {10.3847/1538-4357/ace5aa}, 954, 80

\bibitem[\protect\citeauthoryear{Hagstotz, Reischke  \& Lilow}{Hagstotz et~al.}{2022}]{Hagstotz:2021jzu}
Hagstotz S.,  Reischke R.,   Lilow R.,  2022, \mn@doi [Mon. Not. Roy. Astron. Soc.] {10.1093/mnras/stac077}, 511, 662

\bibitem[\protect\citeauthoryear{{Heintz} et~al.,}{{Heintz} et~al.}{2020}]{Heintz2020}
{Heintz} K.~E.,  et~al., 2020, \mn@doi [\apj] {10.3847/1538-4357/abb6fb}, \href {https://ui.adsabs.harvard.edu/abs/2020ApJ...903..152H} {903, 152}

\bibitem[\protect\citeauthoryear{Ibik et~al.}{Ibik et~al.}{2024}]{Ibik:2023ugl}
Ibik A.~L.,  et~al., 2024, \mn@doi [Astrophys. J.] {10.3847/1538-4357/ad0893}, 961, 99

\bibitem[\protect\citeauthoryear{James et~al.}{James et~al.}{2022}]{James:2022dcx}
James C.~W.,  et~al., 2022, \mn@doi [Mon. Not. Roy. Astron. Soc.] {10.1093/mnras/stac2524}, 516, 4862

\bibitem[\protect\citeauthoryear{Keane et~al.}{Keane et~al.}{2016}]{Keane:2016yyk}
Keane E.~F.,  et~al., 2016, \mn@doi [Nature] {10.1038/nature17140}, 530, 453

\bibitem[\protect\citeauthoryear{Kirsten et~al.}{Kirsten et~al.}{2022}]{Kirsten:2021llv}
Kirsten F.,  et~al., 2022, \mn@doi [Nature] {10.1038/s41586-021-04354-w}, 602, 585

\bibitem[\protect\citeauthoryear{Koch~Ocker, Cordes  \& Chatterjee}{Koch~Ocker et~al.}{2021}]{KochOcker:2021fia}
Koch~Ocker S.,  Cordes J.~M.,   Chatterjee S.,  2021, \mn@doi [Astrophys. J.] {10.3847/1538-4357/abeb6e}, 911, 102

\bibitem[\protect\citeauthoryear{Koch~Ocker et~al.}{Koch~Ocker et~al.}{2022}]{KochOcker:2022ook}
Koch~Ocker S.,  et~al., 2022, \mn@doi [Astrophys. J.] {10.3847/1538-4357/ac6504}, 931, 87

\bibitem[\protect\citeauthoryear{Law et~al.}{Law et~al.}{2020}]{Law:2020cnm}
Law C.~J.,  et~al., 2020, \mn@doi [Astrophys. J.] {10.3847/1538-4357/aba4ac}, 899, 161

\bibitem[\protect\citeauthoryear{Law et~al.}{Law et~al.}{2024}]{Law:2023ibd}
Law C.~J.,  et~al., 2024, \mn@doi [Astrophys. J.] {10.3847/1538-4357/ad3736}, 967, 29

\bibitem[\protect\citeauthoryear{Li \& Zhang}{Li \& Zhang}{2016}]{Li:2016vzg}
Li Y.,  Zhang B.,  2016, arXiv:1603.04825

\bibitem[\protect\citeauthoryear{Li, Gao, Ding, Wang  \& Zhang}{Li et~al.}{2018}]{Li:2017mek}
Li Z.-X.,  Gao H.,  Ding X.-H.,  Wang G.-J.,   Zhang B.,  2018, \mn@doi [Nature Commun.] {10.1038/s41467-018-06303-0}, 9, 3833

\bibitem[\protect\citeauthoryear{Li et~al.}{Li et~al.}{2025}]{Li:2025ckl}
Li Y.,  et~al., 2025, arXiv:2503.04727

\bibitem[\protect\citeauthoryear{Lin \& Sang}{Lin \& Sang}{2021}]{Lin:2021syj}
Lin H.-N.,  Sang Y.,  2021, \mn@doi [Chin. Phys. C] {10.1088/1674-1137/ac2660}, 45, 125101

\bibitem[\protect\citeauthoryear{Lin, Li  \& Tang}{Lin et~al.}{2022}]{Lin:2022afm}
Lin H.-N.,  Li X.,   Tang L.,  2022, \mn@doi [Chin. Phys. C] {10.1088/1674-1137/ac5e92}, 46, 075102

\bibitem[\protect\citeauthoryear{Lin, Tang  \& Zou}{Lin et~al.}{2023}]{Lin:2023jaq}
Lin H.-N.,  Tang L.,   Zou R.,  2023, \mn@doi [Mon. Not. Roy. Astron. Soc.] {10.1093/mnras/stad228}, 520, 1324

\bibitem[\protect\citeauthoryear{Lorimer, Bailes, McLaughlin, Narkevic  \& Crawford}{Lorimer et~al.}{2007}]{Lorimer:2007qn}
Lorimer D.~R.,  Bailes M.,  McLaughlin M.~A.,  Narkevic D.~J.,   Crawford F.,  2007, \mn@doi [Science] {10.1126/science.1147532}, 318, 777

\bibitem[\protect\citeauthoryear{Macquart et~al.}{Macquart et~al.}{2020}]{Macquart:2020lln}
Macquart J.~P.,  et~al., 2020, \mn@doi [Nature] {10.1038/s41586-020-2300-2}, 581, 391

\bibitem[\protect\citeauthoryear{Mahony et~al.}{Mahony et~al.}{2018}]{Mahony:2018ddp}
Mahony E.~K.,  et~al., 2018, \mn@doi [Astrophys. J. Lett.] {10.3847/2041-8213/aae7cb}, 867, L10

\bibitem[\protect\citeauthoryear{Marcote et~al.}{Marcote et~al.}{2020}]{Marcote:2020ljw}
Marcote B.,  et~al., 2020, \mn@doi [Nature] {10.1038/s41586-019-1866-z}, 577, 190

\bibitem[\protect\citeauthoryear{McQuinn}{McQuinn}{2014}]{McQuinn:2013tmc}
McQuinn M.,  2014, \mn@doi [Astrophys. J. Lett.] {10.1088/2041-8205/780/2/L33}, 780, L33

\bibitem[\protect\citeauthoryear{Michilli et~al.}{Michilli et~al.}{2023}]{Michilli:2022bbs}
Michilli D.,  et~al., 2023, \mn@doi [Astrophys. J.] {10.3847/1538-4357/accf89}, 950, 134

\bibitem[\protect\citeauthoryear{Mo, Zhu, Wang, Tang  \& Feng}{Mo et~al.}{2022}]{Mo:2022qxz}
Mo J.-F.,  Zhu W.,  Wang Y.,  Tang L.,   Feng L.-L.,  2022, \mn@doi [Mon. Not. Roy. Astron. Soc.] {10.1093/mnras/stac3104}, 518, 539

\bibitem[\protect\citeauthoryear{Mu\~noz, Kovetz, Dai  \& Kamionkowski}{Mu\~noz et~al.}{2016}]{Munoz:2016tmg}
Mu\~noz J.~B.,  Kovetz E.~D.,  Dai L.,   Kamionkowski M.,  2016, \mn@doi [Phys. Rev. Lett.] {10.1103/PhysRevLett.117.091301}, 117, 091301

\bibitem[\protect\citeauthoryear{Niu et~al.}{Niu et~al.}{2022}]{Niu_2022}
Niu C.~H.,  et~al., 2022, \mn@doi [Nature] {10.1038/s41586-022-04755-5}, 606, 873

\bibitem[\protect\citeauthoryear{Pagano \& Fronenberg}{Pagano \& Fronenberg}{2021}]{Pagano:2021zla}
Pagano M.,  Fronenberg H.,  2021, \mn@doi [Mon. Not. Roy. Astron. Soc.] {10.1093/mnras/stab1438}, 505, 2195

\bibitem[\protect\citeauthoryear{Petroff, Hessels  \& Lorimer}{Petroff et~al.}{2019}]{Petroff:2019tty}
Petroff E.,  Hessels J. W.~T.,   Lorimer D.~R.,  2019, \mn@doi [Astron. Astrophys. Rev.] {10.1007/s00159-019-0116-6}, 27, 4

\bibitem[\protect\citeauthoryear{Platts, Weltman, Walters, Tendulkar, Gordin  \& Kandhai}{Platts et~al.}{2019}]{Platts:2018hiy}
Platts E.,  Weltman A.,  Walters A.,  Tendulkar S.~P.,  Gordin J. E.~B.,   Kandhai S.,  2019, \mn@doi [Phys. Rept.] {10.1016/j.physrep.2019.06.003}, 821, 1

\bibitem[\protect\citeauthoryear{{Prochaska} \& {Zheng}}{{Prochaska} \& {Zheng}}{2019}]{Prochaska:2019mkd}
{Prochaska} J.~X.,  {Zheng} Y.,  2019, \mn@doi [\mnras] {10.1093/mnras/stz261}, \href {https://ui.adsabs.harvard.edu/abs/2019MNRAS.485..648P} {485, 648}

\bibitem[\protect\citeauthoryear{{Prochaska} et~al.,}{{Prochaska} et~al.}{2019}]{Prochaska2019}
{Prochaska} J.~X.,  et~al., 2019, \mn@doi [Science] {10.1126/science.aay0073}, \href {https://ui.adsabs.harvard.edu/abs/2019Sci...366..231P} {366, 231}

\bibitem[\protect\citeauthoryear{Qiang, Deng  \& Wei}{Qiang et~al.}{2020}]{Qiang:2019zrs}
Qiang D.-C.,  Deng H.-K.,   Wei H.,  2020, \mn@doi [Class. Quant. Grav.] {10.1088/1361-6382/ab7f8e}, 37, 185022

\bibitem[\protect\citeauthoryear{Rajwade et~al.}{Rajwade et~al.}{2022}]{Rajwade:2022zkj}
Rajwade K.~M.,  et~al., 2022, \mn@doi [Mon. Not. Roy. Astron. Soc.] {10.1093/mnras/stac1450}, 514, 1961

\bibitem[\protect\citeauthoryear{Rajwade et~al.}{Rajwade et~al.}{2024}]{Rajwade:2024ozu}
Rajwade K.~M.,  et~al., 2024, \mn@doi [Mon. Not. Roy. Astron. Soc.] {10.1093/mnras/stae1652}, 532, 3881

\bibitem[\protect\citeauthoryear{Ravi et~al.}{Ravi et~al.}{2019}]{Ravi:2019alc}
Ravi V.,  et~al., 2019, \mn@doi [Nature] {10.1038/s41586-019-1389-7}, 572, 352

\bibitem[\protect\citeauthoryear{Ravi et~al.}{Ravi et~al.}{2022}]{Ravi:2021kqk}
Ravi V.,  et~al., 2022, \mn@doi [Mon. Not. Roy. Astron. Soc.] {10.1093/mnras/stac465}, 513, 982

\bibitem[\protect\citeauthoryear{Ravi et~al.}{Ravi et~al.}{2023}]{DeepSynopticArrayTeam:2022rbq}
Ravi V.,  et~al., 2023, \mn@doi [Astrophys. J. Lett.] {10.3847/2041-8213/acc4b6}, 949, L3

\bibitem[\protect\citeauthoryear{Ravi et~al.}{Ravi et~al.}{2025}]{Ravi:2023zfl}
Ravi V.,  et~al., 2025, \mn@doi [Astron. J.] {10.3847/1538-3881/adc725}, 169, 330

\bibitem[\protect\citeauthoryear{Ryder et~al.}{Ryder et~al.}{2023}]{Ryder:2022qpg}
Ryder S.~D.,  et~al., 2023, \mn@doi [Science] {10.1126/science.adf2678}, 392, 294

\bibitem[\protect\citeauthoryear{Shah et~al.}{Shah et~al.}{2025}]{Shah:2024ywp}
Shah V.,  et~al., 2025, \mn@doi [Astrophys. J. Lett.] {10.3847/2041-8213/ad9ddc}, 979, L21

\bibitem[\protect\citeauthoryear{Shannon et~al.}{Shannon et~al.}{2025}]{Shannon:2024pbu}
Shannon R.~M.,  et~al., 2025, \mn@doi [Publ. Astron. Soc. Austral.] {10.1017/pasa.2025.8}, 42, e036

\bibitem[\protect\citeauthoryear{Shao \& Zhang}{Shao \& Zhang}{2017}]{Shao:2017tuu}
Shao L.,  Zhang B.,  2017, \mn@doi [Phys. Rev. D] {10.1103/PhysRevD.95.123010}, 95, 123010

\bibitem[\protect\citeauthoryear{Sharma et~al.}{Sharma et~al.}{2023}]{DeepSynopticArrayTeam:2023fxs}
Sharma K.,  et~al., 2023, \mn@doi [Astrophys. J.] {10.3847/1538-4357/accf1d}, 950, 175

\bibitem[\protect\citeauthoryear{Sharma et~al.}{Sharma et~al.}{2024}]{Sharma:2024fsq}
Sharma K.,  et~al., 2024, \mn@doi [Nature] {10.1038/s41586-024-08074-9}, 635, 61

\bibitem[\protect\citeauthoryear{Shull, Smith  \& Danforth}{Shull et~al.}{2012}]{Shull:2011aa}
Shull J.~M.,  Smith B.~D.,   Danforth C.~W.,  2012, \mn@doi [Astrophys. J.] {10.1088/0004-637X/759/1/23}, 759, 23

\bibitem[\protect\citeauthoryear{Tendulkar et~al.}{Tendulkar et~al.}{2017}]{Tendulkar:2017vuq}
Tendulkar S.~P.,  et~al., 2017, \mn@doi [Astrophys. J. Lett.] {10.3847/2041-8213/834/2/L7}, 834, L7

\bibitem[\protect\citeauthoryear{Tian et~al.,}{Tian et~al.}{2024}]{Tian:2024ygd}
Tian J.,  et~al., 2024, \mn@doi [Mon. Not. Roy. Astron. Soc.] {10.1093/mnras/stae2013}, 533, 3174

\bibitem[\protect\citeauthoryear{Tingay \& Kaplan}{Tingay \& Kaplan}{2016}]{Tingay:2016tgf}
Tingay S.~J.,  Kaplan D.~L.,  2016, \mn@doi [Astrophys. J. Lett.] {10.3847/2041-8205/820/2/L31}, 820, L31

\bibitem[\protect\citeauthoryear{Wei, Gao, Wu  \& M\'esz\'aros}{Wei et~al.}{2015}]{Wei:2015hwd}
Wei J.-J.,  Gao H.,  Wu X.-F.,   M\'esz\'aros P.,  2015, \mn@doi [Phys. Rev. Lett.] {10.1103/PhysRevLett.115.261101}, 115, 261101

\bibitem[\protect\citeauthoryear{Williams \& Berger}{Williams \& Berger}{2016}]{Williams:2016zys}
Williams P. K.~G.,  Berger E.,  2016, \mn@doi [Astrophys. J. Lett.] {10.3847/2041-8205/821/2/L22}, 821, L22

\bibitem[\protect\citeauthoryear{Wu \& Wang}{Wu \& Wang}{2024}]{Wu:2024iyu}
Wu Q.,  Wang F.-Y.,  2024, \mn@doi [Chin. Phys. Lett.] {10.1088/0256-307X/41/11/119801}, 41, 119801

\bibitem[\protect\citeauthoryear{Wu et~al.,}{Wu et~al.}{2016}]{Wu:2016brq}
Wu X.-F.,  et~al., 2016, \mn@doi [Astrophys. J. Lett.] {10.3847/2041-8205/822/1/L15}, 822, L15

\bibitem[\protect\citeauthoryear{Wu, Wei, Lan, Gao, Dai  \& M\'esz\'aros}{Wu et~al.}{2017}]{Wu:2017yjl}
Wu X.-F.,  Wei J.-J.,  Lan M.-X.,  Gao H.,  Dai Z.-G.,   M\'esz\'aros P.,  2017, \mn@doi [Phys. Rev. D] {10.1103/PhysRevD.95.103004}, 95, 103004

\bibitem[\protect\citeauthoryear{Wu, Zhang  \& Wang}{Wu et~al.}{2021}]{Wu:2021jyk}
Wu Q.,  Zhang G.~Q.,   Wang F.~Y.,  2021, \mn@doi [arXiv:2108.00581] {10.1093/mnrasl/slac022}

\bibitem[\protect\citeauthoryear{Xiao, Wang  \& Dai}{Xiao et~al.}{2021}]{Xiao:2021omr}
Xiao D.,  Wang F.,   Dai Z.,  2021, \mn@doi [Sci. China Phys. Mech. Astron.] {10.1007/s11433-020-1661-7}, 64, 249501

\bibitem[\protect\citeauthoryear{Xu et~al.}{Xu et~al.}{2022}]{Xu:2021qdn}
Xu H.,  et~al., 2022, \mn@doi [Nature] {10.1038/s41586-022-05493-4}, 609, 685

\bibitem[\protect\citeauthoryear{Yang, Wu  \& Wang}{Yang et~al.}{2022}]{Yang:2022ftm}
Yang K.~B.,  Wu Q.,   Wang F.~Y.,  2022, \mn@doi [Astrophys. J. Lett.] {10.3847/2041-8213/aca145}, 940, L29

\bibitem[\protect\citeauthoryear{Yao, Manchester  \& Wang}{Yao et~al.}{2017}]{Yao_2017msh}
Yao J.~M.,  Manchester R.~N.,   Wang N.,  2017, \mn@doi [The Astrophysical Journal] {10.3847/1538-4357/835/1/29}, 835, 29

\bibitem[\protect\citeauthoryear{Zhang}{Zhang}{2023}]{Zhang_2023}
Zhang B.,  2023, \mn@doi [Rev. Mod. Phys.] {10.1103/RevModPhys.95.035005}, 95, 035005

\bibitem[\protect\citeauthoryear{Zhang \& Zhang}{Zhang \& Zhang}{2025}]{Zhang:2025thh}
Zhang Z.-L.,  Zhang B.,  2025, \mn@doi [Astrophys. J. Lett.] {10.3847/2041-8213/adcc30}, 984, L40

\bibitem[\protect\citeauthoryear{Zhang, Yu, He  \& Wang}{Zhang et~al.}{2020}]{Zhang:2020mgq}
Zhang G.~Q.,  Yu H.,  He J.~H.,   Wang F.~Y.,  2020, \mn@doi [Astrophys. J.] {10.3847/1538-4357/abaa4a}, 900, 170

\bibitem[\protect\citeauthoryear{Zhang, Yan, Li, Zhang  \& Wang}{Zhang et~al.}{2021}]{Zhang:2020xoc}
Zhang Z.~J.,  Yan K.,  Li C.~M.,  Zhang G.~Q.,   Wang F.~Y.,  2021, \mn@doi [Astrophys. J.] {10.3847/1538-4357/abceb9}, 906, 49

\bibitem[\protect\citeauthoryear{Zhang et~al.}{Zhang et~al.}{2024}]{Zhang:2024bar}
Zhang X.,  et~al., 2024, arXiv:2411.17516

\bibitem[\protect\citeauthoryear{Zhao, Zhang, Li, Zhang  \& Zhang}{Zhao et~al.}{2022}]{Zhao:2022yiv}
Zhao Z.-W.,  Zhang J.-G.,  Li Y.,  Zhang J.-F.,   Zhang X.,  2022, arXiv:2212.13433

\makeatother
\end{thebibliography}



\bsp	
\label{lastpage}

\appendix
\begin{onecolumn}

\section{The parameters of 117 well-localized FRBs}
The parameters of 117 well-localized FRBs are summarized in Table \ref{tab:localized_FRBs}. The penultimate column is the inferred ${\rm DM_{host}}$ value based on our Bayesian framework. All the DM values are in units of ${\rm pc~cm^{-3}}$. References:
[1] \cite{Chatterjee:2017dqg},
[2] \cite{Tendulkar:2017vuq},
[3] \cite{Gordon:2023cgw},
[4] \cite{CHIMEFRB:2021srp},
[5] \cite{Mahony:2018ddp},
[6] \cite{Bhandari:2021pvj},
[7] \cite{Michilli:2022bbs},
[8] \cite{Marcote:2020ljw},
[9] \cite{Bannister:2019iju},
[10] \cite{Bhandari:2020oyb},
[11] \cite{Macquart:2020lln},
[12] \cite{Shannon:2024pbu},
[13] \cite{Bhardwaj:2021hgc},
[14] \cite{Prochaska2019}, 
[15] \cite{Bhardwaj:2023vha},
[16] \cite{Ibik:2023ugl},
[17] \cite{Niu_2022},
[18] \cite{Ravi:2019alc},
[19] \cite{Heintz2020},
[20] \cite{Law:2020cnm},
[21] \cite{CHIMEFRB:2023myn},
[22] \cite{Kirsten:2021llv},
[23] \cite{Bhardwaj:2021xaa},
[24] \cite{Connor:2024mjg},
[25] \cite{Rajwade:2022zkj},
[26] \cite{Fong:2021xxj},
[27] \cite{Ravi:2021kqk},
[28] \cite{Bhandari:2022ton},
[29] \cite{James:2022dcx},
[30] \cite{Driessen:2023lxj},
[31] \cite{Caleb:2023atr},
[32] \cite{Cassanelli:2023hvg},
[33] \cite{Sharma:2024fsq}, 
[34] \cite{Law:2023ibd},
[35] \cite{Ravi:2023zfl},
[36] \cite{DeepSynopticArrayTeam:2023fxs},
[37] \cite{Gao:2025fcr},
[38] \cite{Li:2025ckl},
[39] \cite{Ryder:2022qpg},
[40] \cite{Rajwade:2024ozu},
[41] \cite{DeepSynopticArrayTeam:2022rbq},
[42] \cite{Amiri:2025sbi},
[43] \cite{Tian:2024ygd},
[44] \cite{Shah:2024ywp},
[45] \cite{Eftekhari:2024bmi}.

\renewcommand\arraystretch{1.2} 
\begin{longtable}{cccccccccc}
\caption{The parameters of 117 well-localized FRBs.}\label{tab:localized_FRBs}\\	
\hline  
Name & $z$ & ${\rm DM_{obs}}$ & ${\rm DM_{ISM}}$ & $\log(M_*/M_\odot)$ & ${\rm SFR}[M_\odot {\rm yr}^{-1}] $ & Age [Gyr] & ${\rm DM_{host}}$ & Ref. \\
\hline
\endfirsthead
\multicolumn{10}{c}{\textit{\---- Continued from previous page}} \\
\hline  
Name & $z$ & ${\rm DM_{obs}}$ & ${\rm DM_{ISM}}$ & $\log(M_*/M_\odot)$ & ${\rm SFR}[M_\odot {\rm yr}^{-1}] $ & Age [Gyr] & ${\rm DM_{host}}$ & Ref. \\
\hline
\endhead
\hline
\endfoot
\hline
\endlastfoot

FRB20121102A & 0.19273 & 557.0 & 188.4 & $8.14_{-0.10}^{+0.09}$ & $0.05_{-0.01}^{+0.02}$ & $5.71_{-1.26}^{+0.96}$ & $208.66_{-112.34}^{+114.14}$ & [1, 2, 3, 4]\\
FRB20171020A & 0.00867 & 114.10 & 38 & 8.95 & 0.13 & -  & $29.79_{-20.20}^{+25.80}$ & [5] \\
FRB20180301A & 0.3304 & 536 & 152 & $9.64_{-0.11}^{+0.11}$ & $1.91_{-0.55}^{+0.64}$ & $4.34_{-1.29}^{+1.10}$ & $135.52_{-87.22}^{+101.58}$ & [3, 6] \\
FRB20180814A & 0.068 & 189.4 & 87 & $10.78_{-0.18}^{+0.12}$ & $<0.32$ & $7.6_{-3.4}^{+3.3}$  & $36.61_{-26.16}^{+37.95}$ & [4, 7]\\
FRB20180916B & 0.0337 &348.76&199.0& $9.91_{-0.05}^{+0.03}$ & $0.04_{-0.02}^{+0.03}$ & $7.73_{-1.22}^{+0.86}$  & $102.88_{-68.35}^{+85.81}$ & [3, 4, 8]\\
FRB20180924B & 0.3214 & 361.42 & 40.5  & $10.39_{-0.02}^{+0.02}$ & $0.62_{-0.24}^{+0.32}$ & $5.63_{-0.75}^{+0.53}$  & $66.92_{-44.02}^{+50.69}$ & [3, 9, 10, 11, 12] \\
FRB20181030A & 0.00385 & 103.5 & 41.1 & $9.76_{-0.18}^{+0.16}$ & $0.35_{-0.07}^{+0.09}$ & $4.8_{-1.8}^{+1.6}$  & $24.96_{-17.39}^{+23.72}$ & [4, 13]\\
FRB20181112A & 0.4755 & 589.0 & 40 & $9.87_{-0.07}^{+0.07}$ & $1.54_{-0.65}^{+0.99}$ & $3.82_{-0.98}^{+0.84}$  & $203.86_{-100.21}^{+81.46}$ & [3, 10, 11, 12, 14]\\
FRB20181220A & 0.0275 & 209.4 & 126 & $9.86^{+0.14}_{-0.12}$ & $2.9^{+1.6}_{-0.9}$ & $2.7_{-1.6}^{+2.4}$  & $48.84_{-34.49}^{+48.77}$ & [4, 15]\\
FRB20181223C & 0.0302 & 112.5 & 20 &$9.29^{+0.16}_{-0.20}$ & $0.15^{+0.12}_{-0.08}$ & $6.6_{-2.1}^{+0.2}$  & $30.20_{-20.21}^{+24.76}$ & [4, 15]\\
FRB20190102C & 0.2913 & 364.5 & 57 & $9.69_{-0.11}^{+0.09}$ & $0.40_{-0.11}^{+0.31}$ & $4.76_{-1.47}^{+1.02}$  & $74.13_{-47.88}^{+54.44}$ & [3, 10, 11, 12] \\
FRB20190110C & 0.12244 & 221.6 & 37.1 & $10.40_{-0.03}^{+0.02}$ & $0.54_{-0.04}^{+0.04}$ & $2.59_{-1.24}^{+0.80}$  & $59.18_{-35.52}^{+37.60}$ & [4, 16] \\
FRB20190303A & 0.064 & 222.4 & 26 & - & - & -  & $104.53_{-39.15}^{+34.55}$ & [4, 7] \\
FRB20190418A & 0.0713 & 184.5 & 71 & $10.27_{-0.17}^{+0.13}$ & $0.8_{-0.6}^{+1.1}$ & $5.5_{-2.5}^{+2.1}$  & $36.87_{-25.90}^{+35.88}$ & [4, 15] \\
FRB20190425A & 0.0312 & 128.2 & 49 & $10.26_{-0.1}^{+0.09}$ & $1.6_{-0.9}^{+1.5}$ & $6.8_{-2.3}^{+2.0}$  & $26.63_{-18.83}^{+26.62}$ &  [4, 15] \\
FRB20190520B & 0.241 & 1204.7 & 60.5 & $9.08_{-0.09}^{+0.08}$ & $0.04_{-0.02}^{+0.04}$ & $5.27_{-1.32}^{+1.02}$ & $1127.21_{-94.72}^{+65.16}$ & [3, 17] \\
FRB20190523A & 0.660 & 760.8 & 37 & $11.07_{-0.06}^{+0.06}$ & $<1.3$ & -  & $263.07_{-134.96}^{+109.07}$ & [18] \\
FRB20190608B & 0.1178 & 339.5 & 37 & $10.56_{-0.02}^{+0.02}$ & $7.03_{-1.15}^{+1.43}$ & $7.13_{-1.21}^{+0.70}$  & $184.41_{-52.14}^{+42.36}$ & [3, 10, 11, 12] \\
FRB20190611B & 0.378 & 322.2 & 57 & $9.57_{-0.12}^{+0.12}$ & $0.53_{-0.26}^{+0.77}$ & $4.45_{-1.34}^{+0.98}$  & $21.43_{-17.18}^{+30.01}$ & [3, 11, 12, 19] \\
FRB20190614D & 0.6 & 959.2 & 83.5 & - & - & -  & $553.39_{-178.95}^{+126.46}$ & [20] \\
FRB20190711A & 0.522 & 594.6 & 57 & $9.10_{-0.23}^{+0.15}$ & $0.95_{-0.50}^{+0.96}$ & $3.54_{-1.36}^{+0.96}$  & $147.01_{-87.40}^{+85.58}$ & [3, 11, 12, 19]\\
FRB20190714A & 0.2365 & 504.7 & 39 & $10.22_{-0.04}^{0.04}$ & $1.89_{-0.72}^{+1.22}$ & $5.48_{-1.02}^{+0.75}$  & $295.84_{-77.37}^{+55.60}$ & [3, 12, 19] \\
FRB20191001A & 0.234 & 506.92 & 44 & $10.73_{-0.08}^{+0.07}$ & $18.28_{-8.95}^{+17.24}$ & $3.89_{-1.56}^{+1.68}$  & $293.73_{-77.73}^{+56.77}$ &  [3, 12, 19] \\
FRB20191106C & 0.10775 & 332.2 & 25 & $10.65_{-0.13}^{+0.10}$ & $4.75_{-1.27}^{+1.29}$ & -  & $196.36_{-48.84}^{+38.93}$ & [16, 21] \\
FRB20191228A & 0.2430 & 297.5 & 33 & 9.73 & $0.50_{-0.15}^{+0.15}$ & -  & $62.51_{-39.76}^{+43.89}$ & [6, 12] \\
FRB20200120E & 0.0008 & 87.77 & 40.8 & $5.77_{-0.22}^{+0.19}$ & - & -  & $17.98_{-12.34}^{+15.68}$ & [22, 23, 24] \\
FRB20200223B & 0.06024 & 201.8 & 45.6 & $10.75_{-0.08}^{+0.08}$ & $0.59_{-0.04}^{+0.04}$ & $3.37_{-0.73}^{+0.47}$  & $67.11_{-36.77}^{+37.01}$ & [16, 21] \\
FRB20200430A & 0.1608 & 380.1 & 27 & $9.30_{-0.10}^{+0.07}$ & $0.11_{-0.04}^{+0.06}$ & $5.99_{-1.31}^{+0.96}$  & $214.67_{-58.34}^{+44.30}$ & [3, 12, 19] \\
FRB20200906A & 0.3688 & 577.8 & 36 & $10.37_{-0.05}^{+0.05}$ & $4.93_{-2.34}^{+3.46}$ & $4.30_{-1.11}^{+0.86}$  & $290.85_{-100.02}^{+70.41}$ & [3, 6, 12] \\
FRB20201123A & 0.0507 & 433.55 & 251.93 & 11.2 & 0.2 & -  & $131.57_{-87.31}^{+109.44}$ &  [25] \\
FRB20201124A & 0.0979 &413.52& 76 & $10.22_{-0.05}^{+0.05}$ & $2.72_{-1.22}^{+1.65}$ & $6.13_{-1.16}^{+0.95}$  & $231.94_{-62.20}^{+54.11}$ & [3, 24, 26, 27] \\
FRB20210117A & 0.214 & 729.1 & 34 & $8.59_{-0.06}^{+0.05}$ & $0.02_{-0.01}^{+0.01}$ & $5.01_{-1.21}^{+0.95}$  & $587.58_{-78.99}^{+52.87}$ &  [3, 12, 28] \\
FRB20210320C & 0.2797 & 384.8 & 42 & $10.37_{-0.06}^{+0.05} $ & $3.51_{-1.45}^{+2.44}$ & $4.56_{-1.15}^{+0.99}$  & $116.67_{-61.26}^{+55.69}$ & [3, 12, 29] \\
FRB20210405I & 0.066 & 565.17 & 516.1 & 11.25 & $>0.3$ & -  & $153.74_{-110.76}^{+165.47}$ & [30] \\
FRB20210410D & 0.1415 & 578.78 & 56.2 & $9.47_{-0.05}^{+0.05}$ & $0.03_{-0.01}^{+0.03}$ & $6.78_{-1.48}^{+1.02}$  & $416.31_{-66.09}^{+51.53}$ & [3, 31] \\
FRB20210603A & 0.1772 & 500.147 & 40 & $10.93_{-0.04}^{+0.04}$ & $0.24_{-0.06}^{+0.06}$ & $4.32_{-0.75}^{+0.73}$ & $327.00_{-67.56}^{+49.77}$ & [32] \\
FRB20210807D & 0.1293 & 251.9 & 121 & $10.97_{-0.02}^{+0.02}$
 & $0.63_{-0.17}^{+0.18}$ & $8.36_{-1.84}^{+2.25}$  & $43.95_{-31.99}^{+48.21}$ & [3, 12] \\
FRB20211127I & 0.0469 & 234.83 & 43 & $9.48_{-0.02}^{+0.06}$ & $35.83_{-1.46}^{+1.02}$ & $3.85_{-3.65}^{+2.13}$  & $109.37_{-40.89}^{+37.12}$ & [3,12] \\
FRB20211203C & 0.3439 & 636.2 & 63 & $9.76_{-0.09}^{+0.07}$ & $15.91_{-2.98}^{+2.82}$ & $2.47_{-1.25}^{+2.00}$  & $349.81_{-105.38}^{+76.99}$ & [3, 12] \\
FRB20211212A & 0.0707 & 206 & 27 & $10.28_{-0.06}^{+0.05} $ & $0.73_{-0.39}^{+0.62}$ & $5.83_{-1.16}^{+1.05}$  & $82.86_{-37.80}^{+34.64}$ & [3, 12] \\
FRB20220105A & 0.2785 & 583 & 22 & $10.01_{-0.07}^{+0.05}$ & $0.42_{-0.19}^{+0.31}$ & $5.67_{-1.24}^{+0.73}$  & $386.56_{-87.86}^{+57.34}$ & [3, 12] \\
FRB20220204A & 0.4012 & 612.20 & 50.7 & $9.7_{-0.09}^{+0.04}$ &  $4.31_{-0.88}^{+1.40}$ & $4.16_{-0.36}^{+0.32}$  & $287.62_{-107.05}^{+78.75}$ & [24, 33] \\
FRB20220207C & 0.0430 & 262.38 & 76.0 & $9.95_{-0.03}^{+0.03}$ & $0.71_{-0.13}^{+0.12}$ & $6.12_{-0.48}^{+0.45}$  & $105.89_{-49.16}^{+47.44}$ & [24, 33, 34] \\
FRB20220208A & 0.3510 & 437.0 & 101.6 & $10.08_{-0.02}^{+0.02}$ &  $0.68_{-0.08}^{+0.07}$ & $5.21_{-0.15}^{+0.23}$  & $73.82_{-51.78}^{+68.82}$ & [24, 33] \\
FRB20220307B & 0.248123 & 499.27 & 128.2 & $10.14_{-0.04}^{+0.03}$ & $3.17_{-1.03}^{+1.35}$ & $3.64_{-0.43}^{+0.45}$  & $171.45_{-91.24}^{+90.06}$ & [24, 33, 34] \\
FRB20220310F & 0.477958 & 462.24 & 46.3 &  $9.98_{-0.06}^{+0.08}$ & $0.97_{-0.29}^{+0.32}$ & $2.88_{-0.67}^{+0.75}$  & $58.74_{-42.31}^{+56.86}$ & [24, 33, 34] \\
FRB20220319D & 0.0111 & 110.98 & 139.8 & $10.10_{-0.00}^{+0.00}$ & $0.42_{-0.02}^{+0.02}$ & $6.12_{-0.06}^{+0.07}$  & $21.56_{-16.00}^{+26.43}$ & [24, 33, 34, 35] \\
FRB20220330D & 0.3714 & 468.1 & 38.6 & $10.50_{-0.01}^{+0.02}$ & $2.01_{-0.07}^{+0.12}$ & $4.08_{-0.21}^{+0.40}$ & $144.98_{-75.21}^{+65.49}$ & [24, 33] \\
FRB20220418A & 0.622000 & 623.25 & 36.7& $10.26_{-0.02}^{+0.02}$ & $1.31_{-0.13}^{+0.15}$ & $2.59_{-0.29}^{+0.27}$  & $120.29_{-78.35}^{+84.63}$ & [24, 33, 34] \\
FRB20220501C & 0.381 & 449.5 & 31 & - & - & -  & $124.55_{-68.96}^{+62.79}$ &  [12] \\
FRB20220506D & 0.30039 & 396.97 & 84.5 & $10.45_{-0.03}^{+0.01}$ &  $1.20_{-0.17}^{+0.17}$  & $4.77_{-0.16}^{+0.20}$  & $77.46_{-51.84}^{+63.11}$ & [24, 33, 34] \\
FRB20220509G & 0.089400 & 269.53 & 55.6 & $10.7_{-0.01}^{+0.01}$ & $0.25_{-0.04}^{+0.07}$ & $7.75_{-0.18}^{+0.18}$  & $106.41_{-47.84}^{+44.14}$ & [24, 33, 34, 36] \\
FRB20220529A & 0.1839 & 250.2 & 39.95 & $9.43_{-0.13}^{+0.10}$ & 0.13 & - & $49.77_{-32.90}^{+39.07}$ & [37, 38] \\
FRB20220610A & 1.015 & 1458.1 & 31 & 9.7 & 1.7 & -  & $1077.91_{-324.00}^{+200.86}$ & [12, 39] \\
FRB20220717A & 0.36295 &637.34& 118 & - & $0.65_{-0.14}^{+0.14}$ & -  & $259.70_{-115.83}^{+102.52}$ & [40] \\
FRB20220725A & 0.1926 & 290.4 & 31 & - & - & -  & $88.58_{-47.10}^{+44.06}$ &  [12] \\
FRB20220726A & 0.3619 & 686.55 & 89.5 & $10.18_{-0.03}^{+0.04}$ & $0.71_{-0.16}^{+0.19}$ & $5.35_{-0.36}^{+0.52}$  & $364.05_{-116.98}^{+90.97}$ & [24, 33] \\
FRB20220825A & 0.241397 & 651.24 & 78.5 & $10.01_{-0.06}^{+0.06}$ & $7.91_{-0.94}^{+1.26}$ & $5.15_{-0.31}^{+0.32}$  & $419.55_{-93.19}^{+71.49}$ & [24, 33, 34] \\
FRB20220831A & 0.2620 & 1146.25 & 126.7 & - & - & -  & $961.50_{-123.93}^{+98.81}$ & [24] \\
FRB20220912A & 0.0771 & 219.46 & 125.2 & $10.0_{-0.1}^{+0.1}$ & $>0.1$ & -  & $42.21_{-30.61}^{+46.08}$ & [41] \\
FRB20220914A & 0.113900 & 631.28 & 54.7& $9.24_{-0.04}^{+0.08}$ & $0.46_{-0.06}^{+0.05}$ & $4.55_{-0.28}^{+0.29}$  & $488.57_{-60.46}^{+48.23}$ & [24, 33, 34, 36] \\
FRB20220918A & 0.491 & 656.8 & 41 & - & - & -  & $280.06_{-118.32}^{+87.54}$ & [12] \\
FRB20220920A & 0.158239 & 314.99 & 39.9 & $9.87_{-0.01}^{+0.01}$ & $1.62_{-0.21}^{+0.20}$ & $4.27_{-0.29}^{+0.27}$  & $127.98_{-53.36}^{+45.62}$ & [24, 33, 34] \\
FRB20221012A & 0.284669 & 441.08 & 54.4 & $10.96_{-0.02}^{+0.02}$ & $0.18_{-0.07}^{+0.10}$ & $6.74_{-0.21}^{+0.26}$  & $162.82_{-73.99}^{+62.70}$ & [24, 33, 34] \\
FRB20221027A & 0.2290 & 452.50 & 47.2 & $9.47_{-0.04}^{+0.07}$ & $0.56_{-0.09}^{+0.16}$ & $3.37_{-0.16}^{+0.13}$  & $227.58_{-73.38}^{+56.46}$ & [24, 33] \\
FRB20221029A & 0.9750 & 1347.10& 43.9 & $10.59_{-0.10}^{+0.14}$  & $5.21_{-2.52}^{+5.54}$ & $2.93_{-0.17}^{+0.21}$  & $884.03_{-297.21}^{+191.50}$ & [24, 33] \\
FRB20221101B & 0.2395 & 490.70 & 131.2 & $11.21_{-0.02}^{+0.03}$  &  $12.27_{-4.31}^{+4.71}$  & $5.49_{-0.24}^{+0.33}$  & $165.08_{-89.80}^{+90.01}$ & [24, 33] \\
FRB20221106A & 0.2044 & 343.8 & 35 & - & - & -  & $133.75_{-57.83}^{+48.47}$ &  [12] \\
FRB20221113A & 0.2505 & 411.40 & 91.7 & $9.48_{-0.04}^{+0.04}$ & $0.24_{-0.07}^{+0.08}$ & $4.16_{-0.96}^{+0.45}$  & $113.96_{-66.79}^{+69.54}$ & [24, 33] \\
FRB20221116A & 0.2764 & 640.60 & 132.3 & $11.01_{-0.02}^{+0.02}$ & $22.58_{-6.06}^{+6.23}$ & $5.64_{-0.17}^{+0.52}$  & $312.11_{-115.28}^{+100.65}$ & [24, 33] \\
FRB20221219A & 0.5530 & 706.70 & 44.4 & $10.21_{-0.04}^{+0.03}$ & $1.78_{-0.23}^{+0.24}$ & $3.59_{-0.15}^{+0.15}$  & $284.40_{-128.14}^{+97.21}$ & [24, 33] \\
FRB20230124A & 0.0939 & 590.60 & 38.5 & $9.46_{-0.00}^{+0.00}$ & $0.75_{-0.04}^{+0.04}$ & $6.85_{-0.07}^{+0.06}$  & $469.98_{-51.81}^{+41.23}$ & [24, 33] \\
FRB20230203A & 0.1464 & 420.1 & 36.29 & - & - & -  & $257.84_{-59.18}^{+45.35}$ &  [42] \\
FRB20230216A & 0.5310 & 828.0 & 38.5 & $9.82_{-0.01}^{+0.00}$ & $20.81_{-0.17}^{+0.20}$ & $1.27_{-0.03}^{+0.04}$ & $494.22_{-151.54}^{+97.79}$ & [24, 33] \\
FRB20230222A & 0.1223 & 706.1 & 134.13 & - & - & -  & $474.98_{-94.26}^{+83.65}$ & [42] \\
FRB20230222B & 0.1100 & 187.8 & 27.7 & - & - & -  & $45.08_{-28.83}^{+32.75}$ &  [42] \\
FRB20230307A & 0.2706 & 608.90 & 37.6 & $10.76_{-0.02}^{+0.03}$ & $0.46_{-0.06}^{+0.06}$ & $7.07_{-0.40}^{+0.27}$  & $403.42_{-88.60}^{+59.90}$ & [24, 33] \\
FRB20230311A & 0.1918 & 364.3 & 92.39 & - & - & -  & $103.90_{-60.95}^{+64.15}$ &  [42] \\
FRB20230501A & 0.3015 & 532.50 & 125.6 & $10.29_{-0.02}^{+0.00}$ & $4.10_{-0.45}^{+0.60}$ & $0.60_{-0.01}^{+0.01}$  & $173.26_{-94.76}^{+94.21}$ & [24, 33] \\
FRB20230521B & 1.3540 & 1342.90 & 138.8 & - & - & -  & $237.00_{-170.23}^{+221.35}$ & [24] \\
FRB20230526A & 0.1570 & 361.4 & 50 & - & - & -  & $168.44_{-59.50}^{+49.43}$ & [12] \\
FRB20230626A & 0.3270 & 451.20 & 39.2 & $10.44_{-0.04}^{+0.04}$ & $0.98_{-0.24}^{+0.33}$ & $5.41_{-0.36}^{+0.29}$  & $160.33_{-75.21}^{+62.23}$ & [24, 33] \\
FRB20230628A & 0.1270 & 345.15 & 39.1 & $9.29_{-0.03}^{+0.03}$ & $0.14_{-0.03}^{+0.04}$ & $4.82_{-0.50}^{+0.56}$  & $182.50_{-53.84}^{+43.71}$ & [24, 33] \\
FRB20230703A & 0.1184 & 291.3 & 26.97 & - & - & -  & $142.93_{-48.17}^{+39.68}$ & [42] \\
FRB20230708A & 0.105 & 411.51 & 50 & - & - & -  & $255.75_{-55.29}^{+45.36}$ &  [12] \\
FRB20230712A & 0.4525 & 586.96 & 39.2 & $11.13_{-0.01}^{+0.01}$ & $29.91_{-0.79}^{+0.73}$ & $3.85_{-0.06}^{+0.06}$  & $223.65_{-102.19}^{+79.60}$ & [24, 33] \\
FRB20230718A & 0.035 & 477.0 & 396 & - & - & -  & $130.77_{-93.06}^{+134.88}$ & [12] \\
FRB20230730A & 0.2115 & 312.5 & 85.18 & - & - & -  & $57.55_{-39.83}^{+52.26}$ & [42] \\
FRB20230814B & 0.5535 & 696.40 & 104.9 & - & - & -  & $186.37_{-110.60}^{+112.13}$ &  [24] \\
FRB20230902A & 0.3619 & 440.1 & 34 & - & - & -  & $125.40_{-68.29}^{+61.77}$ &  [12] \\
FRB20230926A & 0.0553 & 222.8 & 52.69 & - & - & -  & $82.85_{-41.10}^{+39.74}$ & [42] \\
FRB20231005A & 0.0713 & 189.4 & 33.37 & - & - & -  & $60.77_{-34.03}^{+34.36}$ & [42] \\
FRB20231011A & 0.0783 & 186.3 & 70.36 & - & - & -  & $36.01_{-25.42}^{+35.56}$ & [42] \\
FRB20231017A & 0.2450 & 344.2 & 64.55 & - & - & -  & $77.11_{-48.70}^{+54.34}$ & [42] \\
FRB20231025B & 0.3238 & 368.7 & 48.67 & - & - & -  & $65.60_{-43.85}^{+52.11}$ & [42] \\
FRB20231120A & 0.0368 & 438.90 & 43.8 & $10.40_{-0.00}^{+0.00}$ & $0.40_{-0.03}^{+0.02}$ & $7.54_{-0.08}^{+0.14}$  & $324.86_{-41.49}^{+36.90}$ & [24, 33] \\
FRB20231123A & 0.0729 & 302.1 & 89.76 & - & - & -  & $114.75_{-56.64}^{+55.33}$ &  [42] \\ 
FRB20231123B & 0.2621 & 396.70 & 40.2 & $11.04_{-0.01}^{+0.01}$ & $4.85_{-0.21}^{+0.22}$ & $4.86_{-0.09}^{+0.10}$  & $145.55_{-66.16}^{+55.47}$ & [24, 33] \\
FRB20231128A & 0.1079 & 331.6 & 25.05 & - & - & -  & $195.56_{-48.85}^{+38.94}$ &  [42] \\ 
FRB20231201A & 0.1119 & 169.4 & 70.03 & - & - & -  & $22.49_{-16.81}^{+27.49}$ &  [42] \\ 
FRB20231204A & 0.0644 & 221.0 & 29.73 & - & - & -  & $98.89_{-39.38}^{+35.18}$ & [42] \\
FRB20231206A & 0.0659 & 457.7 & 59.13 & - & - & -  & $314.87_{-51.64}^{+44.75}$ &  [42] \\
FRB20231220A & 0.3355 & 491.20 & 49.9 & - & - & -  & $188.52_{-83.20}^{+67.57}$ &  [24] \\
FRB20231223C & 0.1059 & 165.8 & 47.9& - & - & -  & $25.21_{-18.34}^{+27.39}$ &  [42] \\
FRB20231226A & 0.1569 & 329.9 & 145 & - & - & -  & $68.77_{-48.44}^{+67.05}$ & [12] \\
FRB20231229A & 0.0190 & 198.5 & 58.12 & - & - & -  & $75.39_{-38.39}^{+38.26}$ &  [42] \\
FRB20231230A & 0.0298 & 131.4 & 61.51 & - & - & -  & $25.68_{-18.47}^{+27.54}$ &  [42] \\
FRB20240114A & 0.1300 & 527.65 &49.7& - & - & -  & $371.96_{-61.46}^{+48.04}$ & [43] \\
FRB20240119A & 0.3700 & 483.1 & 37.9 & - & - & -  & $165.24_{-80.24}^{+66.62}$ &  [24] \\
FRB20240123A & 0.9680 & 1462.00 & 90.3 & - & - & -  & $1016.88_{-312.24}^{+208.49}$ & [24] \\
FRB20240201A & 0.042729 & 374.5 & 38 & - & - & -  & $261.71_{-40.97}^{+35.92}$ &  [12] \\
FRB20240209A & 0.1384 & 176.49 & 55.5 & $11.35_{-0.01}^{+0.01}$ & $<0.31$ & $11.34_{-2.34}^{+0.03}$  & $21.25_{-15.94}^{+25.87}$ & [44, 45] \\ 
FRB20240210A & 0.023686 & 283.73 & 31 & - & - & -  & $185.62_{-35.76}^{+32.35}$ & [12] \\
FRB20240213A & 0.1185 & 357.40 & 40.1 & - & - & -  & $200.06_{-53.53}^{+43.36}$ & [24] \\
FRB20240215A & 0.2100 & 549.50 & 48.0 & - & - & -  & $355.43_{-76.49}^{+55.82}$ & [24] \\
FRB20240229A & 0.2870 & 491.15 & 37.9 & - & - & -  & $243.52_{-82.04}^{+60.26}$ & [24] \\
FRB20240310A & 0.1270 & 601.8 & 36 & - & - & -  & $472.98_{-58.32}^{+43.90}$ & [12] \\
\end{longtable}

\end{onecolumn}

\end{document}